\documentclass[conference]{IEEEtran}

\usepackage{tikz}

\usepackage{amsmath}
\usepackage{indentfirst}
\usepackage{epigraph} 
\usepackage{booktabs}
\usepackage{array}
\usepackage{hyperref}
\usepackage{mdframed}
\usepackage{utfsym}
\usepackage{xurl}
\usepackage[misc]{ifsym}

\usepackage{bm}

\def\ie{$i.e.$}

\usepackage{color,colortbl,array,xspace}
\usepackage{enumitem}

\hypersetup{
  colorlinks   = true,    
  urlcolor     = blue,    
  linkcolor    = blue,    
  citecolor    = blue      
}

\newcommand{\sys}{\text{Prometheus}\xspace}

\newcommand{\para}[1]{\vspace{2pt}\noindent{\textbf{#1}}\vspace{0.1pt}}

\definecolor{lightgray}{gray}{0.9}

\newenvironment{icompact}{
  \begin{list}{$\bullet$}{
    \itemindent -.05em
    \parsep 0pt plus 1pt
    \partopsep 0pt plus 1pt
    \topsep 2pt plus 2pt minus 2pt
    \itemsep 0pt plus 1.3pt
    \parskip 0pt plus 2pt
    \leftmargin 0.13in}
      }
{\normalsize
\end{list}
}

\usepackage[many]{tcolorbox}

\definecolor{darkgrey}{HTML}{434343}

\definecolor{revblue}{RGB}{0,114,178}
\newcommand{\rev}[1]{\textcolor{revblue}{#1}}

\newtcolorbox{policybox}[2][]{text width=0.95\linewidth,fontupper=\normalsize,
fonttitle=\bfseries\sffamily\scriptsize, colbacktitle=darkgrey,enhanced,
attach boxed title to top left={yshift=-2mm,xshift=3mm},
boxed title style={sharp corners},top=4pt,bottom=2pt,left=2pt,right=2pt,
  title=#2,colback=white}

\usepackage{listings}
\usepackage{hyperref}
\hypersetup{
    colorlinks=true,
    linkcolor=blue,
    filecolor=magenta,      
    urlcolor=cyan,
    pdftitle={Overleaf Example},
    pdfpagemode=FullScreen,
    }

\usepackage{microtype}
\usepackage{booktabs} 
\usepackage{multirow}  %
\usepackage{utfsym} 
\usepackage{bbm} 
\usepackage{algpseudocode}
\usepackage{algorithm} 
\usepackage{subfigure}
\usepackage{pifont}
\usepackage{hyperref}
\usepackage{ulem} 

\begin{document}

\title{Towards Effective Prompt Stealing Attack against\\Text-to-Image Diffusion Models}


\author{
\IEEEauthorblockN{
Shiqian Zhao\IEEEauthorrefmark{2},
Chong Wang\IEEEauthorrefmark{2}$ ^{\textrm{\Letter}}$,
Yiming Li\IEEEauthorrefmark{2},
Yihao Huang\IEEEauthorrefmark{3},
Wenjie Qu\IEEEauthorrefmark{3},
Siew-Kei Lam\IEEEauthorrefmark{2},
Yi Xie\IEEEauthorrefmark{4},\\
Kangjie Chen\IEEEauthorrefmark{2},
Jie Zhang\IEEEauthorrefmark{8}$ ^{\textrm{\Letter}}$,
Tianwei Zhang\IEEEauthorrefmark{2}}
\IEEEauthorblockA{\IEEEauthorrefmark{2}Nanyang Technological University, Singapore}
\IEEEauthorblockA{\IEEEauthorrefmark{3}National University of Singapore, Singapore}
\IEEEauthorblockA{\IEEEauthorrefmark{4}Tsinghua University, China}
\IEEEauthorblockA{\IEEEauthorrefmark{8}Agency for Science, Technology and Research (A*STAR), Singapore}
}

\IEEEoverridecommandlockouts
\makeatletter\def\@IEEEpubidpullup{4.5\baselineskip}\makeatother
\IEEEpubid{\parbox{\columnwidth}{
\rule{\columnwidth/2}{0.5pt}\\
$^{\textrm{\Letter}}$ Corresponding authors: Chong Wang \textless chong.wang@ntu.edu.sg\textgreater, Jie Zhang \textless zhangj6@a-star.edu.sg\textgreater \\ \\
Network and Distributed System Security (NDSS) Symposium 2026\\
23 - 27 February 2026, San Diego, CA, USA\\
ISBN 979-8-9919276-8-0\\  
https://dx.doi.org/10.14722/ndss.2026.241899\\
www.ndss-symposium.org
}
\hspace{\columnsep}\makebox[\columnwidth]{}}

\maketitle
\begin{abstract}
  
Text-to-Image (T2I) models, represented by DALL$\cdot$E and Midjourney, have gained huge popularity for creating realistic images. The quality of these images relies on the carefully engineered prompts, which have become valuable intellectual property. While skilled prompters showcase their AI-generated art on markets to attract buyers, this business incidentally exposes them to \textit{prompt stealing attacks}. Existing state-of-the-art attack techniques reconstruct the prompts from a fixed set of modifiers (\textit{i.e.,} style descriptions) with model-specific training, which exhibit restricted adaptability and effectiveness to diverse showcases (\textit{i.e.,} target images) and diffusion models. 

To alleviate these limitations, we propose \textbf{Prometheus}, a training-free, proxy-in-the-loop, search-based prompt-stealing attack, which reverse-engineers the valuable prompts of the showcases by interacting with a local proxy model. It consists of three innovative designs. 
First, we introduce \textit{dynamic modifiers}, as a supplement to static modifiers used in prior works. These dynamic modifiers provide more details specific to the showcases, and we exploit NLP analysis to generate them on the fly. 
Second, we design a \textit{contextual matching} algorithm to sort both dynamic and static modifiers. This offline process helps reduce the search space of the subsequent step. 
Third, we interact with a local proxy model to invert the prompts with a greedy search algorithm. Based on the feedback guidance, we refine the prompt to achieve higher fidelity. 
The evaluation results show that \textbf{Prometheus} successfully extracts prompts from popular platforms like PromptBase and AIFrog against diverse victim models, including Midjourney, Leonardo.ai, and DALL$\cdot$E, with an ASR improvement of 25.0\%. 
We also validate that \textbf{Prometheus} is resistant to extensive potential defenses, further highlighting its severity in practice. Our code is available at \url{https://github.com/Shiqian-Zhao996/Prometheus}.
  
\end{abstract}
\section{Introduction}
\begin{figure}[t!]
  \centering
  \includegraphics[width=\linewidth]{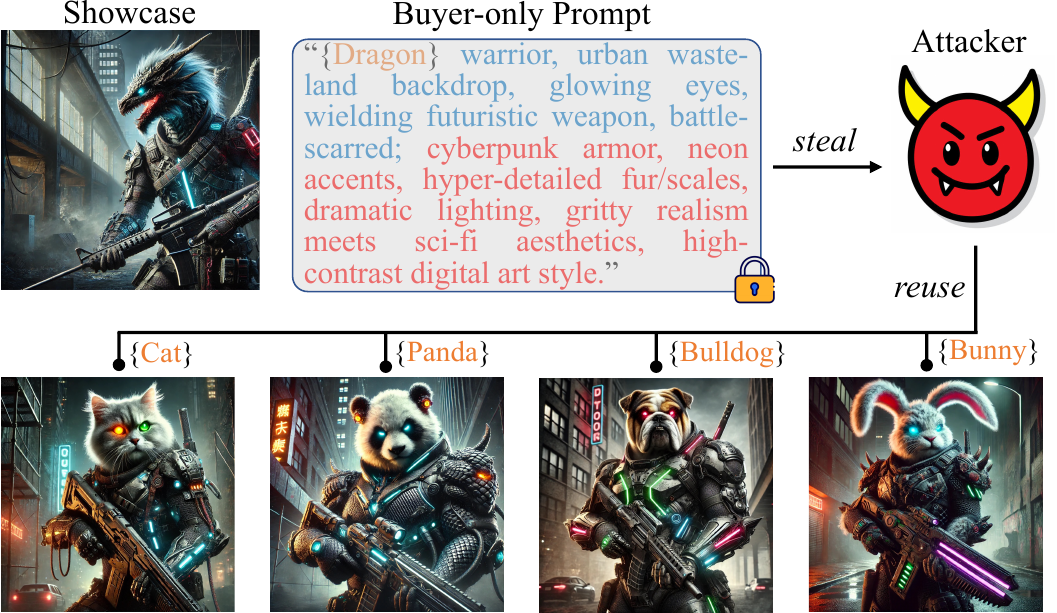} 
  \vspace{-0.1in}
  \caption{A prompt and its corresponding images. The blue phrase is the subject, and the red phrases are the modifiers. After the attacker steals the prompt, they can reuse it by replacing the subject (in brown). This prompt is from the commercial prompt market PromptBase~\cite{PromptBase}, whose showcase is generated with DALL$\cdot$E 3. }
  \vspace{-0.1in}
  \label{fig: showcase}
\end{figure}

With the emergence of diffusion models~\cite{diffusion_1, diffusion_2}, text-to-image generation services, such as DALL$\cdot$E~\cite{DallE} and Midjourney~\cite{Midjourney}, have received widespread popularity due to their remarkable performance. 
Since diffusion models are often sensitive to user inputs, crafting high-quality images requires carefully designed prompts, which typically consist of precise \textit{subjects} (main object) and diverse \textit{modifiers} (style descriptions)~\cite{PromptContent1, PromptContent2, PromptContent3} as shown in Figure~\ref{fig: showcase}. 
The high demand for high-quality prompts has led to the rise of numerous trading platforms, such as PromptBase~\cite{PromptBase}, PromptSea~\cite{PromptSea}, and PromptHero~\cite{PromptHero}, where prompt engineers publish their high-quality prompts, and users purchase their desired prompts by browsing the corresponding showcase images. For example, PromptBase hit 10K registered users by November 2022~\cite{PromptBaseUser}, only 4 months after its establishment in July 2022~\cite{PromptBaseFounding}. \textit{After purchasing the prompt, the buyer can modify and customize it to generate tailored outputs, as well as scale the quantity of images to suit various augmentation needs.}

Prompts sold on the trading platforms often command high prices (\textit{e.g.}, 15 US dollars for the Portrait Neon Lights prompt from Promptrr~\cite{PromptExample}). Thus, they can be considered a form of valuable intellectual property (IP). 
Their great commercial value spawns potential threats to IP protection, and one prominent example is prompt stealing attack~\cite{PromptStealing}, which has already attracted attention from the industry~\cite{WhitePaper}. 
As shown in Figure~\ref{fig: showcase}, given a showcase that is exhibited on the prompt market, an attacker tries to recover the corresponding prompt used by the publisher to create the showcase. \textit{Then the attacker can adapt the prompt to his own tasks, including customization or scaling quantity.} For example, he can reuse the prompt to generate more images by simply replacing the subjects with customized ones. 
Despite the infringement on intellectual property, the attacker can subsequently upload the stolen prompt to other marketplaces for sale. These significantly compromise the commercial interests of both platforms and creators. 

The essence of prompt-stealing attacks lies in accurately extracting the subjects and modifiers from a given showcase image. However, research on these attacks has significantly lagged behind their threat. We summarize existing prompt recovery techniques in Table~\ref{tab: taxonomy}, where BLIP, CLIP-IG, and PH2P mostly focus on the subject, while PromptStealer~\cite{PromptStealing} represents the only dedicated prompt-stealing attack that fully takes into account modifiers. 
PromptStealer takes the first dedicated step by (i) training a caption model for subject generation and (ii) training a classification model for modifier prediction. 
Although PromptStealer has demonstrated promising results, the following limitations reduce its practical penetration effect. 
First, PromptStealer relies on a fixed, predefined set of 7,672 modifiers as classification labels. This approach restricts its semantic coverage, resulting in \textit{out-of-vocabulary (OOV) issues} for modifiers and limiting its ability to describe the highly diverse showcases. 
Second, the predictions for subject and modifiers rely entirely on the caption and classification models, which are trained with the prompt-side text semantics while ignoring the perception and semantics from the recovered images. This may lead to prediction error, especially when the trained models are \textit{overfitted to the training set} (\textit{i.e.}, the Lexica dataset~\cite{Lexica} collected from Stable Diffusion~\cite{stable_diffusion}).

To alleviate these limitations, we propose \sys, an effective and practical prompt-stealing attack methodology. 
Our key insights are twofold. First, we propose \textbf{dynamic modifiers} that are specific to the target showcase, serving as the complement of the predefined but insufficient static modifiers used in prior works. Second, to address the overfitting caused by text-guided pretraining, we incorporate a \textbf{training-free proxy-in-the-loop} mechanism with feedback to iteratively refine the modifiers. 
Specifically, \sys is designed to include three key modules. (1) \textit{Dynamic modifier extraction}. \sys first leverages a caption model to generate lots of image captions for the showcase, which provides abundant showcase-specific modifiers. We then analyze these captions with an NLP analysis tool (\textit{e.g.,} Spacy~\cite{spacy}), to extract these modifiers from the captions. (2) \textit{Contextual matching}. These dynamic modifiers, along with the predefined static modifiers, are concatenated to the subject and ranked based on our proposed contextual matching, which considers the fidelity gain of a modifier in a contextual manner. This action helps to shorten the list of candidate modifiers and further improves the efficiency of subsequent fine-grained sampling. (3) \textit{Greedy $\texttt{Proxy}$ query}. Finally, \sys greedily constructs the final prompt by sequentially adding modifiers to the base prompt (starting with the subject and gradually expanding) and leverages the feedback from a local proxy model to carefully choose the modifiers with the most gain effect. We construct a \textit{multi-objective score function}, \textit{i.e.,} \textit{semantic} and \textit{perception}, to comprehensively evaluate the contribution of a modifier. 

We conduct comprehensive experiments to assess \sys. 
First, large-scale case studies reveal that it effectively steals prompts that are designed for popular commercial platforms like Midjourney~\cite{Midjourney}, and DALL$\cdot$E~\cite{DallE}, from prompt markets like PromptBase~\cite{PromptBase} and AIFrog~\cite{AIFrog}, highlighting its \textit{functionality} in practical scenarios. 
For instance, on real-world prompts, \sys achieves image and prompt fidelity scores of 0.912 and 0.814, respectively, surpassing the baselines by margins of up to 0.142 and 0.200. 
Importantly, our method realizes an attack success rate improvement of 25.0\% compared with the state-of-the-art. 
Moreover, \sys achieves the best \textit{reusability} when shifting the subject in the recovered prompt. 
Further mitigation experiments show that our attack is \textit{resistant to various potential defenses}, including random noise, puzzle effect, text watermark, and adaptive mitigation, highlighting its practicality and severity. 

\begin{table}[t!]
\centering
\caption{Existing and potential prompt-stealing attacks. }
\vspace{-0.1in}
\label{tab: taxonomy}
\resizebox*{\linewidth}{!}{
\begin{tabular}{cccc}
\toprule
\textbf{Method} & \textbf{Static Modifier} & \textbf{Dynamic Modifier} & \textbf{Feedback} \\ \midrule
BLIP~\cite{BLIP} &$\usym{2718}$&$\usym{2718}$&$\usym{2718}$ \\
CLIP-IG~\cite{Interrogator}&$\usym{2714}$&$\usym{2718}$&$\usym{2718}$ \\
PH2P~\cite{mahajan2024prompting} &$\usym{2718}$&$\usym{2718}$&$\usym{2718}$ \\
PromptStealer~\cite{PromptStealing}&$\usym{2714}$&$\usym{2718}$&$\usym{2718}$ \\ 
\sys (Ours)&$\usym{2714}$&$\usym{2714}$&$\usym{2714}$ \\ \bottomrule
\end{tabular}
}
\vspace{-0.1in}
\end{table}

In summary, our main contributions are as follows:
\begin{icompact}
    \item We review existing prompt-stealing attacks against text-to-image models and reveal their poor adaptability. We identify that this limitation arises from the over-preparation. 
    \item We propose a novel and effective training-free prompt-stealing attack, dubbed \sys. It generates modifiers on the fly and leverages the feedback from \texttt{Proxy} to optimize the prompt effectively. 
    \item We propose contextual matching, which is inspired by in-context learning. This mechanism helps sort out the high-correlation modifier to facilitate prompt stealing. 
    \item We evaluate \sys on extensive prompts collected from commercial prompt markets, \textit{e.g.,} PromptBase and AIFrog. The result shows that \sys can steal prompts with high fidelity and is resistant to potential defenses. 
\end{icompact}

\para{Responsible Disclosure:} 
We obtained prior consent from both the prompt engineers and the platforms for the scientific use of the displayed prompt showcases and the collected data. They have expressed great interest in and support for our work. Upon acceptance, we will share the research findings with them, along with mitigation methods for such attacks. 
We will open-source the code and collected dataset for community use. 

\section{Background and Related Work}

\subsection{Text-to-image Models}
Text-to-image (T2I) models~\cite{diffusion_1, diffusion_2} are a new emerging technology for high-quality image generation with text description as a condition, also known as \textit{prompt}. 
During the generation process, a diffusion model is commonly adopted, \textit{e.g.}, U-Net~\cite{UNet-2} to recover the desirable image from random noise, with the guidance of prompts. 
Specifically, these models take the text embedding of the prompt from a fixed text encoder (usually the text encoder of CLIP~\cite{CLIP}). 
Then this embedding is fed into the diffusion model to predict the step-wise noise with the cross-attention mechanism~\cite{Attention, attention-2}. 
Popular T2I models include commercial ones like DALL$\cdot$E~\cite{DallE}, Midjourney~\cite{Midjourney}, Imagen~\cite{Imagen} and open-source Stable Diffusion~\cite{stable_diffusion}. 

\subsection{Prompt-as-a-Service} 
Despite the remarkable generation ability of T2I models, performing prompt engineering requires a significant amount of effort and cost. For instance, PromptPerfect charges up to \$100 per month for its prompt optimization service~\cite{PromptPerfect}. Under these circumstances, Prompt-as-a-Service (PaaS) has emerged as a popular application. Professional prompt engineers optimize prompts with their expertise and sell them on the prompt markets, \textit{e.g.}, PromptBase~\cite{PromptBase}, AIFrog~\cite{AIFrog}, PromptSea~\cite{PromptSea}, PromptHero~\cite{PromptHero}. These on-sell prompts are sightlessly exhibited with their exquisite showcases, which visually indicate the prompt effects. After purchasing, the user gains access to its corresponding textual description, which can be used to generate customized images or increase the number of showcased outputs.

\subsection{Prompt Stealing Attacks} 
Several methods can be potentially used to investigate the feasibility of prompt-stealing attacks (PSA). A straightforward strategy is image captioning via a caption model. For example, Li et al.~\cite{BLIP} train an image-to-text model, BLIP, to predict the prompt of an image. However, as shown in~\cite{PromptContent1, PromptContent2, PromptContent3}, a high-quality prompt should consist of a subject and modifiers. For this reason, the open-source project CLIP-IG~\cite{Interrogator} considers adding modifiers into the subject by combining highly correlated phases from five kinds of modifiers (\textit{i.e.,} medium, artist, trending, movement, and flavor). 
In a more related work, Shen et al.~\cite{PromptStealing} propose to utilize a multi-head model ML-Decoder~\cite{MLDecoder} to detect the modifier contained in the showcase. To train such a predictor, they collect a dataset called Lexica-Dataset. 
Some approaches attempt to reverse-engineer prompts using token-level optimization techniques~\cite{wen2023hard, mahajan2024prompting}, where a soft prompt is first derived and then projected into a hard prompt. However, the resulting prompts often lack semantic coherence, limiting their reusability. Additionally, these methods typically assume white-box access to the generation model, an unrealistic assumption in closed-source settings. More analysis can be found in Section~\ref{sec: main result}. 

Despite their progress in promoting the performance of PSA, these works suffer from two defects. First, they lack feedback from text-to-image models. This means that they rely on a predefined pattern, which may lead to an overfitting problem and thus poor reusability. Second, a lot of manual labor is required in the loop, including collecting the dataset and training additional models, which significantly affects the attack efficiency and cost. 

\section{Problem Statement}

\subsection{Threat Model}
\label{sec: definition}
We follow the same threat model in \cite{PromptStealing}, as detailed below. 

\para{Attacker's Capability.} 
We make two assumptions about the attacker's capabilities. (1) The attacker can access the generated showcases of the prompt (denoted as $s$). This is typical in commercial platforms, where prompt sellers display showcase images to attract potential buyers. However, the specific prompt used to generate the showcase remains hidden from all users until they purchase it. A detailed example can be found in Portrait Neon Lights~\cite{PromptExample}. (2) As some commercial victim models run in a pay-as-you-go mode, which may be expensive, we assume the attacker can use a substitute local text-to-image (T2I) model, referred to as the proxy model. Since the proxy model can be any open-source model, the query budget is 0.

\para{Adversary's Motivation.} 
The attacker aims to compromise the confidentiality of valuable prompts sold on the markets. A carefully designed prompt consists of \textit{subjects} and \textit{modifiers} (as shown in Figure~\ref{fig: showcase}). Specifically, subjects are the main body of a prompt that outlines an image's most important elements, while modifiers detail the style of the subjects and the whole picture. Formally, given a showcase image $s$, an adversary aims to recover the prompt $p$ that is used to create the showcase, including both its subjects and modifiers. We provide more analysis about the motivation in Appendix~\ref{appendix: motivation}. 

From a \textbf{utility perspective}, stolen prompts can be reused for customization, \textit{e.g.,} replacing subjects or increasing the number of showcases. From a \textbf{commercial standpoint}, as discussed in previous work \cite{PromptStealing}, the motivation behind prompt stealing can be twofold. First, it may \textit{infringe on the intellectual property (IP) of high-quality prompts}, which is the result of significant effort by prompt engineers. Second, attackers have strong \textit{economic incentives}: they can avoid payment by using these premium prompts for free and may also redistribute or resell the prompts on other markets for additional profit. This poses a significant threat to the commercial interests of both prompt developers and the platforms hosting them, which is explicitly prohibited by platform PromptBase~\cite{userterms}. From a \textbf{privacy protection standpoint}, this type of attack poses \textit{a significant threat to the confidentiality of prompts, particularly when they contain sensitive or highly classified information.} We provide more analysis for this standpoint in Appendix~\ref{appendix: privacy protection standpoint}.

\subsection{Attack Requirements}
\label{sec:attack-req}
For the attack to be considered successful, the attacker must meet the following requirements: 
\begin{icompact}
    \item \textbf{Objective-1: Functionality [Output Perception Similarity]}. The stolen prompt should achieve a high reproduction in image appearance to the showcase. 
    \item \textbf{Objective-2: Reusability [Subject Shift]}. The recovered prompt possesses high reusability so that when swapping the main subject, it still generates images with similar appearances. 
\end{icompact}

\begin{figure*}[ht!]
  \centering
  \includegraphics[width=0.9\linewidth]{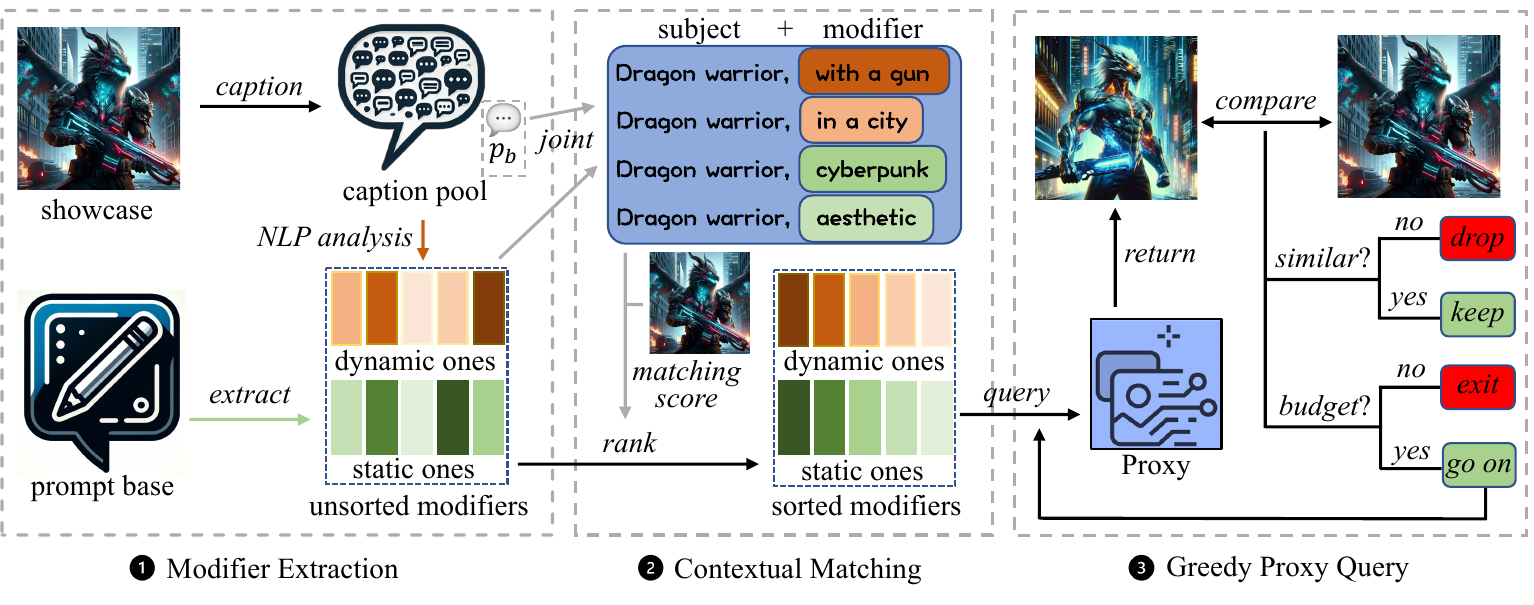} 
  \scriptsize
  \vspace{-0.2in}
  \caption{Overall pipeline of \sys. In general, \sys consists of three main components: \ding{182} Modifier Extraction; \ding{183} Contextual Matching; and \ding{184} Greedy \texttt{Proxy} Query. \sys starts with a showcase and public prompt base, which provides static and dynamic modifiers. Then, contextual matching is utilized to rank and shortlist the unsorted modifiers. \sys interactively queries a local \texttt{Proxy} model and computes the fidelity gain for each modifier (as defined in Equation~\ref{eq:fidelity}). This gain serves as feedback to guide the refinement of the modifiers.}
  \vspace{-0.1in}
  \label{fig: scheme}
\end{figure*}

\para{Attack Comparisons.}
We follow the existing line of work~\cite{BLIP, Interrogator, PromptStealing}, and focus on reversing readable prompts, which could satisfy these two objectives. We observe that certain hard prompt recovery methods could serve as potential solutions. For instance, PH2P~\cite{mahajan2024prompting} leverages gradient signals to optimize a soft prompt, which is then projected back into a hard prompt. However, since semantic coherence is not taken into account during optimization, the recovered prompts often fall into a local optimum and consist of unreadable or nonsensical words. Another drawback of hard prompt recovery methods is that they search modifier candidates within the entire discrete token space, which is rather inefficient. For example, when recovering a 40-word prompt from a 50,000-token vocabulary, PH2P would consider $50,000^{40}$ combinations, which is computationally infeasible. As a result, they fail to meet the aforementioned objectives, limiting the practical utility of the extracted prompts. We list the merits of our \sys over the existing works in Table~\ref{tab: taxonomy}. Besides static modifiers, we also include \textit{dynamic modifiers} and \textit{feedback from the proxy model}, to steal prompts with higher fidelity from a meaningful modifier set efficiently. 

\section{\sys} \label{sec:method}

\subsection{Overview}
\label{sec: overview}

We introduce \sys, an advanced prompt-stealing attack to overcome the limitations of existing works. 
\begin{icompact}
    \item \textbf{Showcase-specific Dynamic Modifiers.} In addition to the fixed static modifiers used in the previous work \cite{PromptStealing}, we propose to generate a pool of \textit{dynamic modifiers} for target showcases on the fly, to provide more comprehensive descriptions. 
    \item \textbf{Proxy-in-the-loop Feedback.} We propose to leverage \texttt{Proxy}'s feedback to optimize the prompt for the desired effect, \textit{i.e.,} showcase. The attacker queries the \texttt{Proxy} with a combined modifier pool. Based on the \texttt{Proxy}'s feedback, \textit{i.e.,} each modifier's fidelity gain to the showcase $s$, the attacker decides to keep or discard the modifier. 

\end{icompact}

Figure \ref{fig: scheme} depicts the workflow of \sys. Given a showcase, it employs a captioning model (\textit{e.g.}, BLIP) to generate a caption as the initial subject. It then strategically enhances this subject by adding a set of modifiers to create the final prompt. Specifically, \sys consists of the following three key steps to achieve the requirements of \textit{functionality} and \textit{reusability} as stated in Section \ref{sec:attack-req}. 

\para{1. Modifier Extraction} (Section \ref{sec:modifier-extract}).
First, we construct a comprehensive and semantic modifier pool to ensure functionality and reusability. 
We satisfy these requirements by considering two kinds of modifiers: 1) \textit{Static modifiers}, which are extracted from a luxuriant prompt base in prior work~\cite{PromptStealing}, comprising prompt engineers' prior knowledge; 2) \textit{Dynamic modifiers}, which are extracted from the target showcase's captions, and generated on the fly. They serve as a supplement to fill in more showcase-specific details dynamically. 

\para{2. Contextual Matching} (Section \ref{sec:in-context}).
After obtaining the comprehensive modifier list, it is inefficient to query all of these modifiers to select the desired ones. Therefore, we \textit{refine them to narrow down the search space}. Considering the modifiers' decorative effect on the subject, we propose a \textit{contextual matching} algorithm to refine these modifiers for streamlining the sample space. We verify that this intuitive matching method could achieve a more consistent ranking of the modifiers. 

\begin{figure*}[ht!]
  \centering
  \includegraphics[width=0.9\linewidth]{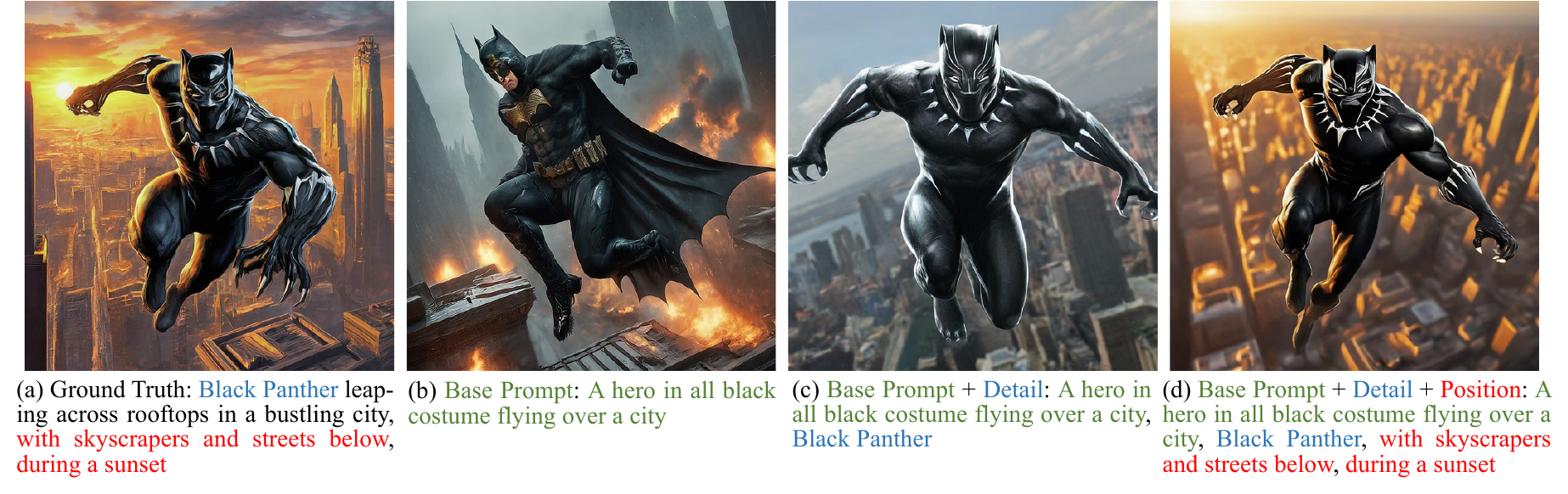} 
  \scriptsize
  \vspace{-0.10in}
  \caption{An illustration of dynamic modifiers. The base prompts in $(b)$–$(d)$ are generated by the caption model BLIP~\cite{BLIP} based on the showcase. As shown, the caption model fails to capture detailed subject and position information. Additionally, in the transition from $(a)$ to $(b)$, these omissions result in an image that differs from the showcase. However, this issue can be mitigated by incorporating detailed subject and position information, as demonstrated in $(c)$ and $(d)$. The evaluation model used is Imagen~\cite{Imagen} from Gemini. }
  \vspace{-0.1in}
  \label{fig: dynamic modifier}
\end{figure*} 

\begin{figure}[ht!]
  \centering
  \includegraphics[width=0.9\linewidth]{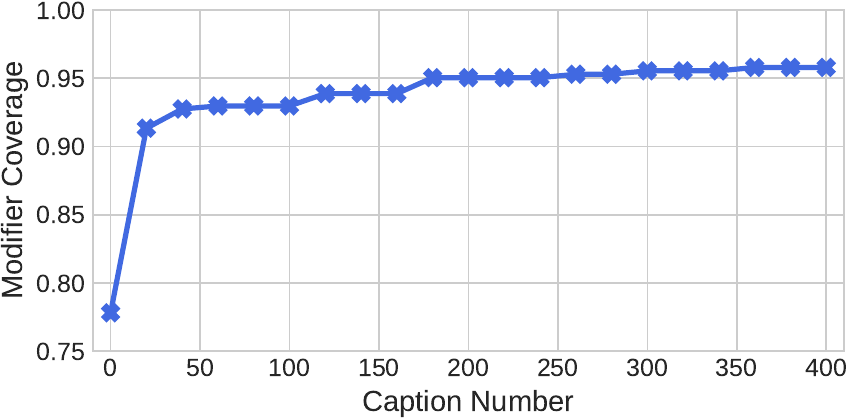} 
  \caption{Modifier coverage with caption number. }
  \label{fig: modifier coverage}
  \vspace{-0.1in}
\end{figure}

\para{3. Greedy \texttt{Proxy} Query} (Section \ref{sec:greed-query}).
Third, we optimize the prompts to achieve higher functionality. Since the local ranking from contextual matching may not fully align with the target, it is crucial to perform additional refinement based on feedback from the \texttt{Proxy}. We employ the \textit{greedy search algorithm} to enhance the modifiers further. Specifically, for each prompt candidate, we query the \texttt{Proxy} and compute the similarity between its output and the showcase. Using the similarity gain as a criterion, we decide whether to retain or discard each modifier. This greedy approach allows us to identify the most effective modifiers. The entire interaction completes within $\mathcal{Q}$ turns. 

Below we provide details of each step in \sys. 

\subsection{Modifier Extraction}
\label{sec:modifier-extract}
Previous works~\cite{Interrogator, PromptStealing} primarily focus on adopting the overall style of a showcase as the prompt modifier. These are referred to as \textit{static modifiers} because they are extracted from a fixed set of prompts and remain largely unchanged across images. Typically, they rely on the captioning model BLIP~\cite{BLIP} to extract the main subjects of showcases. However, as illustrated in Figure~\ref{fig: dynamic modifier}, such captioning models often fail to capture the details of a given showcase $s$ with sufficient accuracy~\cite{PromptStealing}, which may lead to \textit{unmatched image generation}. For example, BLIP describes ``Black Panther'' merely as a ``hero in all black costume'', which can easily be interpreted as ``Batman'' by a text-to-image model. Also, \textit{spatial cues} such as ``during a sunset'' can offer vital information about scene layout, yet are often missed. 

To address this, we introduce two additional image elements, including \textit{subject detail} and \textit{position information}, to enrich the image description. In this paper, these are collectively referred to as \textbf{dynamic modifiers}, as they vary with each showcase. Formally, the \textbf{dynamic modifiers} are defined as the subject details (\textit{e.g.}, identity and specified property) and the position information (\textit{i.e.}, the whole layout of showcase), which are beyond the static style modifiers. 

\para{Dynamic Modifier.} We present the pseudocode (Appendix~\ref{appendix: dynamic modifier extraction}) along with the following introduction to detail dynamic modifier extraction. We leverage the zero-shot capability of caption model to extract dynamic modifiers. Specifically, given a showcase, we first generate multiple prompts with a local BLIP model. This caption model could generate captions with large variances under a high sampling temperature, finding all the potential descriptions. Then, for each sampled caption, we utilize sentence analysis techniques to extract \textit{nominal and prepositional phrases}. Specifically, we utilize Spacy~\cite{spacy} to analyze the prompt for \textit{Parts of Speech (PoS)} and construct a parent-child tree for each word. 
For subject detail, we extract all the noun chunks in the prompt, including single nouns (\textit{e.g.}, ``Black Panther'') and noun phrases (\textit{e.g.}, ``red car''). 
For position information, the extraction process is much more complicated. First, \sys goes through one caption to locate all the preposition words. If one word is a preposition, then a prepositional phrase is constructed based on its relation tree. 
Take the phrase ``during a sunset'' as an example: when \sys detects that ``during'' is a preposition, the relation tree of ``during'' is extracted, \textit{e.g.,} ``a'' is the child of ``sunset'' and ``a sunset'' is the parent of ``during''. 
Finally, we obtain a prepositional phrase by appending the child to its corresponding parent. 
We do not adopt an object detector as it can only recognize the category of a subject (\textit{e.g.,} car) while \textit{ignoring fine-grained details} (\textit{e.g.,} the color of a car). 
Also, it cannot get the \textit{layout} of an image, \textit{i.e.,} position information.

We validate the effectiveness of dynamic modifiers in Figure~\ref{fig: modifier coverage}. As illustrated, incorporating dynamic modifiers extracted from the caption pool significantly enhances modifier coverage. Specifically, static modifiers alone account for 77.8\% coverage. When dynamic modifiers are introduced, coverage increases to 95.8\% as the caption pool expands. This improvement demonstrates that \textit{dynamic modifiers effectively compensate for the limitations of both the BLIP model and static modifiers.}

\subsection{Contextual Matching}
\label{sec:in-context}
After obtaining the modifiers, \sys ranks them by their correlation to the showcase. This ranking process occurs locally to narrow the search space for more cost-effective \texttt{Proxy} queries. 
An intuitive way is to apply CLIP to calculate the modifiers' semantic similarity to the showcase~\cite{Interrogator}. 
However, we argue that, as modifiers are appendants to a base prompt, modifier correlation alone does not indicate the gain to prompt fidelity. Therefore, we have to rethink the effect contextually. Driven by this, we propose a new correlation ranking method, namely \textit{Contextual Matching}, to measure a modifier's fidelity gain to the subject comprehensively. 

\begin{figure}[t!]
  \centering
  \includegraphics[width=0.9\linewidth]{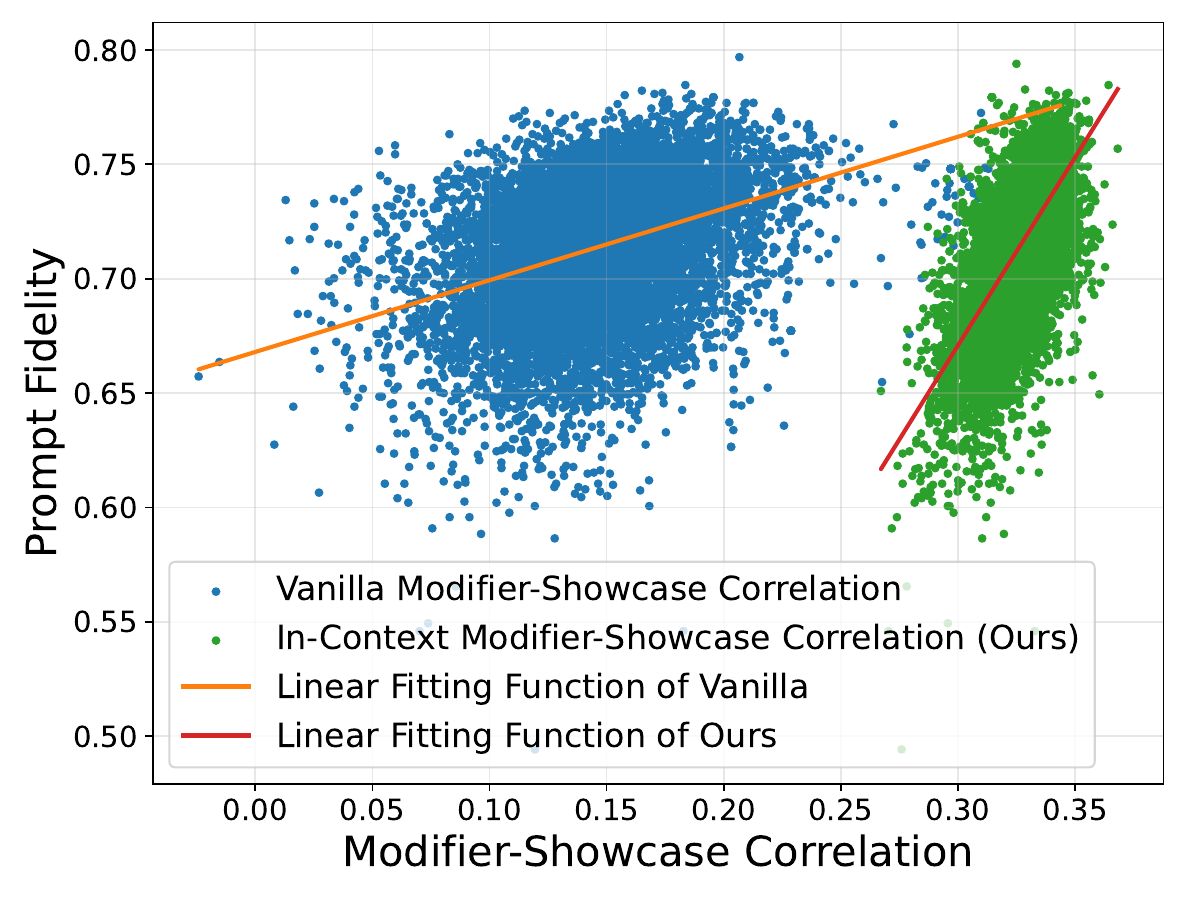} 
  \caption{Comparison between our contextual modifier ranking method and the baseline. The slopes of fitted linear functions are the PCC. }
  \label{fig: in-context rank}
  \vspace{-0.1in}
\end{figure}

Specifically, given a showcase $s$ and a base prompt $p_b$, we calculate the contextual matching score of modifier $m$ with the following score function $\mathcal{F}_{score}$:
\begin{equation}
\label{eq:score function}
    \mathcal{F}_{score} = \mathcal{D}(\mathcal{E}_{T}(p_b+m), \mathcal{E}_{I}(s))-\mathcal{D}(\mathcal{E}_{T}(p_b), \mathcal{E}_{I}(s)), 
\end{equation}
where $\mathcal{E}_{T}$ and $\mathcal{E}_{I}$ are the text encoder and image encoder of CLIP, respectively. The distance function $\mathcal{D}$ is cosine similarity. As CLIP is trained to build a correlation between texts and images, this metric could reflect the matching degree of a modifier in context. 

We validate the superiority of our contextual modifier ranking method using \textit{Pearson Correlation}. 
We calculate the semantic similarity between contextual modifiers and showcase as $\mathcal{D}(\mathcal{E}_{T}(p_b+m),\mathcal{E}_{I}(s))$, as well as the prompt similarity between $\hat{p}$ and ground truth $p$ as $\mathcal{D}(\mathcal{E}_{T}(p),\mathcal{E}_{T}(\hat{p}))$. Then we calculate the Pearson Correlation Coefficient (PCC) between these two similarities. We compare our contextual matching with the vanilla matching method~\cite{Interrogator}, which directly calculates the similarity between the modifier and showcase. 
As shown in Figure~\ref{fig: in-context rank}, our contextual matching presents \textit{a more positive correlation}. More specifically, our method achieves a PCC of 0.65, while the vanilla method only achieves 0.37, indicating that \textit{our method has a more consistent trend with the victim}. We present a more detailed ablation study in Section \ref{sec: eval}.

\subsection{Greedy \texttt{Proxy} Query}
\label{sec:greed-query}

From the first two steps, we obtain two lists of modifier candidates for the dynamic and static ones, respectively. 
These candidates are generated, ranked, and refined using BLIP~\cite{BLIP} and CLIP~\cite{CLIP}. However, due to the knowledge gap between these models and the T2I generative model, it is not practical to directly select the most probable modifiers from these lists (truncation). 
Simply choosing modifiers based on relevance ranking is insufficient, as the achieved PCC of 0.65 by contextual matching is still inadequate, though it already outperforms the baseline ranking method. 
We argue that truncating the top-$k$ modifiers risks under- or overslicing, which can lead to suboptimal recovery performance. To address this, we propose leveraging feedback from a local proxy model to \textit{greedily search} for modifiers that provide the largest fidelity gain. Specifically, the attacker $\mathcal{A}$ interacts with the \texttt{Proxy} by iteratively extending the base prompt, adding one modifier at a time. Modifiers are then selected based on the \texttt{Proxy}'s feedback, \textit{i.e.,} only those that enhance the quality of the final prompt $\hat{p}$ are retained. 
This iterative refinement process ensures maximum fidelity between the recovered prompt and the ground-truth prompt.

\para{Score Function.} To accurately assess the fidelity gain of a modifier, a comprehensive scoring function is essential. Intuitively, the perception of similarity between the recovered image (\textit{i.e.,} the image generated using the stolen prompt) and the showcase can serve as a useful guide. However, because the showcase includes many random elements introduced by the inherent randomness of any T2I model, \textit{relying too heavily on perception consistency risks severe overfitting}. Specifically, the stolen prompt may end up including additional modifiers beyond the ground truth to account for this randomness. To address this, we leverage \textit{both semantic and perceptual guidance} to refine the modifiers. Let $p_{b}$ represent the base prompt, which is updated to $p_q$ with modifier $m_q$, and let the corresponding feedback from the \texttt{Proxy} $\mathcal{O}$ be denoted as $\mathcal{O}(p_{b})$ and $\mathcal{O}(p_q)$, respectively. The fidelity gain is:
\begin{equation}\label{eq:fidelity}
    \Delta (m_{q}) = Sem(p_{b}, p_{q}, s) + Per(p_{b}, p_{q}, s). 
\end{equation}
Here, $fn = Sim(\mathcal{O}(p_q), s) - Sim(\mathcal{O}(p_{b}), s)$, where $fn \in \left\{Sem, Per\right\}$, and $Sim$ represent the corresponding similarity metrics. Specifically, the similarity function for image semantics is the CLIP score~\cite{CLIP}, and that for image perception is LPIPS~\cite{LPIPS}. Note that the base prompt is dynamic, that is, when the fidelity gain $\Delta (m_{q})$ meets the threshold $\delta$, the base prompt is updated as $p_b+m_q$. Then it is regarded as the base prompt for the next modifier. 

\renewcommand{\algorithmicrequire}{\textbf{Input:}}
\renewcommand{\algorithmicensure}{\textbf{Output:}}
\algnewcommand{\Cmnt}[1]{\Comment{\textnormal{\textcolor{gray}{\small\em #1}}}}
\begin{algorithm}[!t] \footnotesize
    \caption{$\mathtt{GreedyProxyQuery}$}
    \label{algo-1}
    \begin{algorithmic}[1] 
        \Require base prompt $p_b$, showcase $s$, sorted modifier set $\mathcal{M}$, fidelity gain threshold $\delta$, proxy model $\mathcal{O}$, and allocated query budget $\mathcal{Q}$. 
        \Ensure recovered prompt $\hat{p}$.
    \State $s_b$ $\gets$ $\mathcal{O}(p_b)$ \Cmnt{Query the proxy model}
    \For{$q$ in $range(\mathcal{Q})$}
        \State $p_q$ $\gets$ joint $\mathcal{M}[q]$ to $p_b$ \Cmnt{Joint candidate to base prompt}
        \State $s_q$ $\gets$ $\mathcal{O}(p_q)$ \Cmnt{Query the proxy model}
        \State $\Delta (\mathcal{M}[q]) = Sem(s_{b}, s_{q}, s) + Per(s_{b}, s_{q}, s)$
        \If{$\Delta (\mathcal{M}[q])>\delta$} \Cmnt{If the threshold is reached}
            \State $p_b=p_q$
            \State $s_b=s_q$ \Cmnt{Update base prompt and best score}
        \EndIf
    \EndFor
    \State $\hat{p}$ $\gets$ assign $p_b$ as final prompt
    \State \textbf{return} $\hat{p}$
    \end{algorithmic}
    \label{alg:greedysearch}
\end{algorithm}

\begin{table*}[ht!]
\centering
\renewcommand{\arraystretch}{1.2} 
\setlength{\tabcolsep}{3.5pt}
\scriptsize
\caption{Performance of \sys compared with baseline methods. The models referenced here represent the proxy models used for prompt extraction. RealPrompt consists of prompts sold on commercial markets (${CLIP}_{img}:\uparrow$, $LPIPS:\downarrow$, $SBERT:\uparrow$, $ASR:\uparrow$). }
\vspace{-0.1in}
\label{tab: main result}
\resizebox*{\linewidth}{!}{
\begin{tabular}{cc|cccc|cccc|cccc}
\toprule
\multirow{2}{*}{\textbf{Dataset}} & \multirow{2}{*}{\textbf{Method}} & \multicolumn{4}{c|}{\textbf{FLUX}} & \multicolumn{4}{c|}{\textbf{ShuffleDiffusion}} & \multicolumn{4}{c}{\textbf{Stable Diffusion-3.5}} \\
&  &$\texttt{CLIP}_{img}$&LPIPS&SBERT&ASR&$\texttt{CLIP}_{img}$&LPIPS&SBERT&ASR&$\texttt{CLIP}_{img}$&LPIPS&SBERT&ASR   \cr
\midrule
\multirow{6}{*}{DALLEPrompt}&BLIP&0.694&0.776&0.721&24.7\%&0.746&0.733&0.747&26.0\%&0.729&0.773&0.734&15.3\% \\
&CLIP-IG&0.798&0.745&0.729&23.3\%&0.828&0.701&0.733&22.7\%&0.822&0.723&0.736&14.7\% \\
&PromptStealer&0.731&0.761&0.733&13.3\%&0.767&0.725&0.734&28.7\%&0.778&0.757&0.741&18.7\% \\
&PH2P&0.542&0.836&0.531&0\%&0.490&0.771&0.612&0\%&0.656&0.751&0.599&0\% \\
&VLMasExpert&0.785&0.726&0.746&26.7\%&0.814&0.674&0.767&38.7\%&0.818&0.703&0.757&29.3\% \\
&\cellcolor{lightgray}Ours&\cellcolor{lightgray}\textbf{0.873}&\cellcolor{lightgray}\textbf{0.679}&\cellcolor{lightgray}\textbf{0.795}&\cellcolor{lightgray}\textbf{48.3\%}&\cellcolor{lightgray}\textbf{0.890}&\cellcolor{lightgray}\textbf{0.601}&\cellcolor{lightgray}\textbf{0.803}&\cellcolor{lightgray}\textbf{56.0\%}&\cellcolor{lightgray}\textbf{0.897}&\cellcolor{lightgray}\textbf{0.626}&\cellcolor{lightgray}\textbf{0.790}&\cellcolor{lightgray}\textbf{52.7\%} \cr \midrule \midrule
\multirow{6}{*}{RealPrompt}&BLIP&0.768&0.757&0.724&14.6\%&0.745&0.740&0.672&8.3\%&0.758&0.761&0.710&16.7\% \\
&CLIP-IG&0.842&0.708&0.755&29.2\%&0.848&0.712&0.775&35.4\%&0.851&0.708&0.778&27.1\% \\
&PromptStealer&0.771&0.731&0.765&35.4\%&0.770&0.743&0.780&31.3\%&0.781&0.762&0.781&37.5\% \\ 
&PH2P&0.760&0.759&0.617&0\%&0.815&0.719&0.598&0\%&0.793&0.795&0.611&0\% \\
&VLMasExpert&0.848&0.695&0.776&27.1\%&0.858&0.688&0.748&29.2\%&0.854&0.685&0.774&33.3\% \\ 
&\cellcolor{lightgray}Ours &\cellcolor{lightgray}\textbf{0.901}&\cellcolor{lightgray}\textbf{0.653}&\cellcolor{lightgray}\textbf{0.814}&\cellcolor{lightgray}\textbf{58.3\%}&\cellcolor{lightgray}\textbf{0.912}&\cellcolor{lightgray}\textbf{0.644}&\cellcolor{lightgray}\textbf{0.798}&\cellcolor{lightgray}\textbf{54.2\%}&\cellcolor{lightgray}\textbf{0.901}&\cellcolor{lightgray}\textbf{0.625}&\cellcolor{lightgray}\textbf{0.799}&\cellcolor{lightgray}\textbf{62.5\%} \cr
\bottomrule
\end{tabular}
}
\end{table*}

\para{Greedy Proxy Query.} We present our $\mathtt{GreedyProxyQuery}$ process in Algorithm~\ref{alg:greedysearch}, which contains four main steps:
\begin{enumerate}[noitemsep,topsep=0pt,parsep=0pt,partopsep=0pt,leftmargin=*]
    \item Sample a modifier $m$ as the order provided in Section~\ref{sec:in-context} and append it to the base prompt $p_b$;
    \item Query $p_b$ with the proxy model $\mathcal{O}$ and obtain feedback $\eta$ as score function shown in Equation~\ref{eq:score function};
    \item If $\eta$ does not meet threshold $\delta$, discard modifier $m$; otherwise, append $m$ to $p_b$ and obtain the new $p_b$;
    \item Repeat Steps (1)-(3) until the budget is exhausted or the prompt fidelity requirement is met. 
\end{enumerate}

As dynamic modifiers serve as a supplement to the subject, we begin the searching process by examining the dynamic modifier list ($m_d$) before proceeding to the static modifier list ($m_s$). To ensure balanced consideration of both dynamic and static modifiers, we pre-allocate the budget between $m_d$ and $m_s$ in a 1:4 ratio. The final optimized prompt is constructed by combining the subject with both dynamic and static modifiers, as formulated below: 
\begin{equation}
    \hat{p} = p_b + m_d + m_s. 
\end{equation}

\section{Evaluation} 
\label{sec: eval}

\subsection{Experiment Setup}

\para{Datasets.} We consider two datasets: DALLEPrompt and RealPrompt. Each data point in these sets consists of a showcase and the corresponding prompt. 
\begin{icompact}
    \item \textbf{DALLEPrompt.} This dataset comprises prompts sourced from real-world prompts designed for DALL$\cdot$E~\cite{dalleprompt}. It spans various styles, including oil paintings, pixel art, and cyberpunk, as well as themes such as movie posters, book covers, and early 1900s newspapers. The prompts encompass a wide array of subjects, featuring characters (\textit{e.g.}, Mickey Mouse, Darth Vader), objects (\textit{e.g.}, marbles, boomboxes), and locations (\textit{e.g.}, castles, Ancient Egypt). We generated showcases with these prompts using three well-performing models: FLUX~\cite{flux}, ShuttleDiffusion~\cite{shuttlediffusion}, and Stable Diffusion-3.5~\cite{sd3.5} (SD-3.5). Finally, we obtained three sets containing prompt-showcase pairs (50 pairs for each). 
    \item \textbf{RealPrompt.} This collection comprises a meticulously curated set of 24 prompt-showcase pairs, sourced from real-world commercial prompt markets including AIFrog~\cite{AIFrog} and PromptBase~\cite{PromptBase}. These pairs contain eight distinct themes, each showcasing a unique style. For every style, three images were generated by the sellers using prompts featuring different subjects. The average word prompt length is 32.08, with an average of 9.11 modifiers. The associated generation models include Midjourney~\cite{Midjourney} (1 set), Stable Diffusion~\cite{stable_diffusion} (3 sets), Leonardo.ai~\cite{Leonardo} (2 sets), and DALL$\cdot$E~\cite{dalle2} (2 sets). 
\end{icompact}

\para{Baselines.} We consider the only prompt stealing method and four prompt recovery methods as baselines. The content in brackets denotes the original task. 

\begin{figure*}[ht!]
  \centering
  \includegraphics[width=0.9\linewidth]{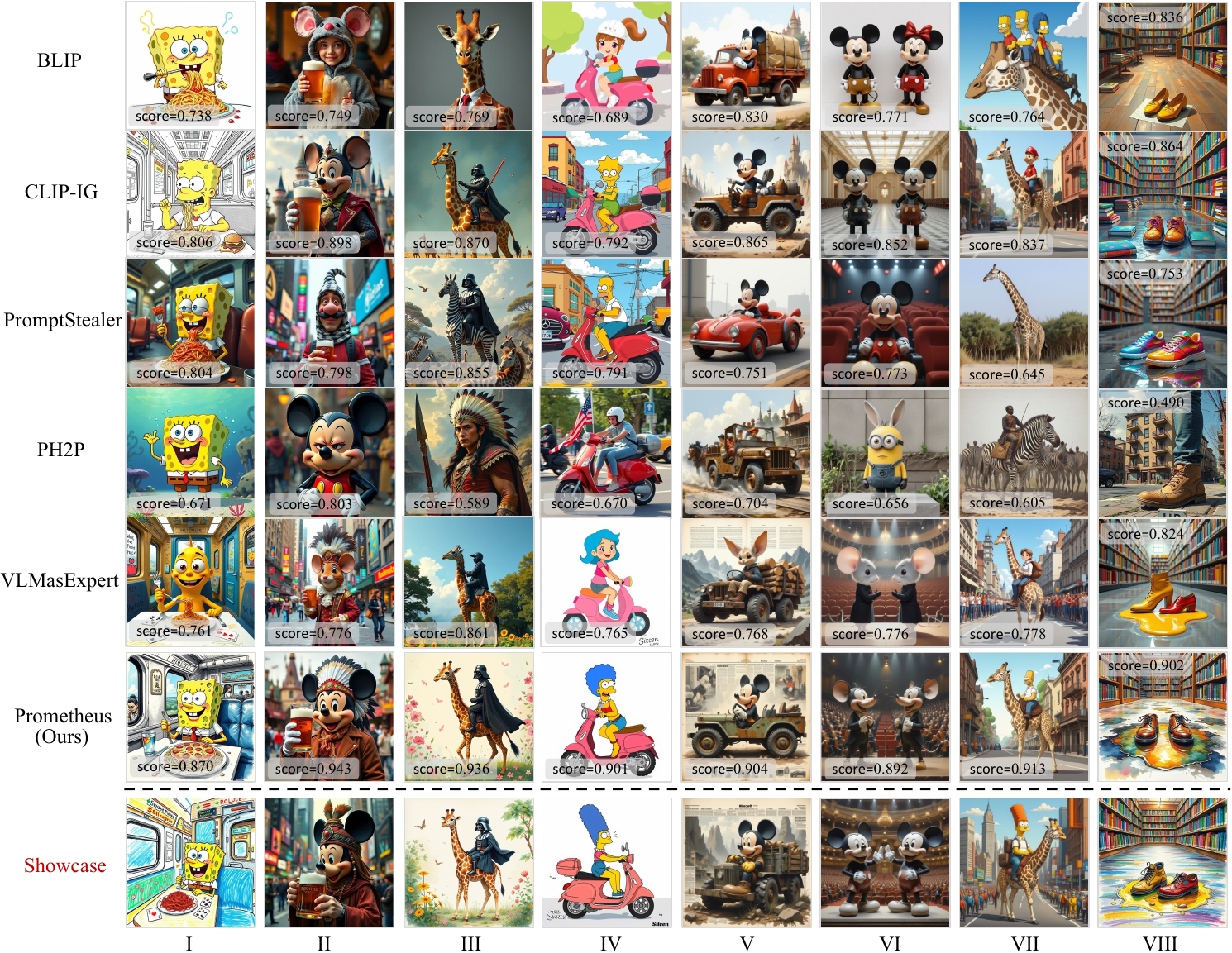} 
  \scriptsize
  \caption{Visualization of generated images compared with showcase. We provide eight different themes here. An example of the corresponding prompts can be found in Table~\ref{tab: example} (case-VI). The \textit{score} on the generations represents the CLIP score ($\uparrow$), which indicates the semantic similarity between the generation and the showcase. Please zoom out for details. }
  \label{fig: visualization-1}
  \vspace{-0.1in}
\end{figure*}

\begin{icompact}
    \item \textbf{BLIP} (\textit{Captioning})~\cite{BLIP}: This is a captioning model trained on image-caption pairs from the COCO dataset~\cite{MSCOCO}. Since the captions primarily focus on describing the subjects in an image, BLIP often exhibits suboptimal performance in capturing styles or modifiers. 
    \item \textbf{CLIP-IG} (\textit{Captioning})~\cite{Interrogator}: This method builds upon BLIP by enhancing its capabilities. In addition to the subjects generated by BLIP, CLIP-IG selects a set of modifiers from a predefined large-scale modifier pool. These modifiers are ranked based on their semantic similarity to the showcase, with text-to-image similarity assessed using CLIP's encoders. 
    \item \textbf{PromptStealer} (\textit{Prompt Stealing})~\cite{PromptStealing}: This method is specifically designed for prompt stealing against text-to-image models. Given a showcase, it utilizes a fine-tuned BLIP model to generate the subject and employs a fine-tuned multi-head classifier to predict modifiers from a predefined modifier pool. The fine-tuning process relies on the Lexica prompt-image dataset, which was collected using the Stable Diffusion models. 
    \item \textbf{PH2P} (\textit{Prompt Inversion})~\cite{mahajan2024prompting}: We also consider PH2P, a method that inverts the prompt with access to the encoder of Stable Diffusion v1.5. PH2P first performs token-level optimization on the soft prompt using gradients, and then projects the optimized soft prompt into a hard prompt. 
    \item \textbf{VLMasExpert} (\textit{VLM}): We include an additional baseline, Vision-LLM as an expert (VLMasExpert), in our study. Vision-LLM, trained on extensive datasets for image analysis, can be likened to a human expert with specialized knowledge. We utilize the powerful GPT-4o~\cite{hurst2024gpt} to generate descriptions of the showcases. 

\end{icompact}

\para{Evaluation Metrics.} 
We consider four metrics to evaluate the stolen prompt, including: 
\begin{icompact}
\item $\textbf{CLIP}_{img}$: This metric evaluates the semantic similarity between generated image $g$ and showcase $s$. It is calculated with image encoder $\mathcal{E}_{img}$ from CLIP~\cite{CLIP} as:
\begin{equation}
    {CLIP}_{img}(g,s)=\frac{\mathcal{E}_{img}(g)\cdot\mathcal{E}_{img}(s)}{\|\mathcal{E}_{img}(g)\|\|\mathcal{E}_{img}(s)\|}, \nonumber
\end{equation}
where a higher score indicates greater semantic similarity.

\item \textbf{LPIPS}: The Learned Perceptual Image Patch Similarity (LPIPS) metric is designed to align closely with human perception~\cite{LPIPS}. It evaluates the visual similarity between two images using features extracted by neural network (\textit{e.g.}, AlexNet backbone). We use LPIPS to assess the perceptual similarity between the recovered image and showcase. A lower LPIPS indicates better perception alignment. 

\item \textbf{SBERT}: SentenceBERT (SBERT) is an enhanced version of BERT~\cite{reimers2019sentence}, designed specifically for measuring sentence-level similarity. We use this metric to evaluate the semantic similarity between the stolen prompt $\hat{p}$ and the victim prompt $p$. 
Specifically, SBERT score functions as follows:
\begin{equation}
{SBERT}(\hat{p},p)=\frac{\mathcal{E}_{bert}(\hat{p})\cdot\mathcal{E}_{bert}(p)}{\|\mathcal{E}_{bert}(\hat{p})\|\|\mathcal{E}_{bert}(p)\|},
    \nonumber
\end{equation}
where $\mathcal{E}_{bert}$ is the embedding extractor of SentenceBert. 
\item \textbf{ASR}: We also use the Attack Success Rate (ASR) as a metric to evaluate the effectiveness of attacks. Following the setting in~\cite{li2018textbugger, garg2020bae, roth2024constraint}, an attack is considered successful if the SBERT score is greater than 0.8 (\textit{e.g.}, the semantics of $\hat{p}$ and $p$ are similar enough).
\end{icompact}

We agree that no single metric fully captures similarity or usefulness, and selecting the top image by one metric can conceal weaknesses on others. Accordingly, for the three attacks, we do not pick a single ``best'' test by any one metric; instead, we report the average across all metrics. Besides these metrics, we also conducted a large-scale user study, which directly evaluates the human perception of the stolen effect. The detailed experimental design and results are presented in Section~\ref{sec: user study}. 

\begin{table}[!t]
\centering
\renewcommand{\arraystretch}{1.3} 
\setlength{\tabcolsep}{0.7pt}
\scriptsize
\caption{Examples of target and stolen prompts generated by \sys and baseline methods. The corresponding images are shown in Figure~\ref{fig: transfer-1}. The Blue annotations indicate the parts that match the target prompt. }
\vspace{-0.1in}
\label{tab: example}
\begin{tabular}{lm{6.7cm}}
\toprule
\textbf{Target prompt} & Two Mickey Mice are talking in a concert hall; sculpture, hyperrealistic. \cr 
\midrule \midrule
\textbf{BLIP} & \textcolor{blue}{Two figurines} are positioned to \textcolor{blue}{Mickey Mouse}. \cr \midrule
\textbf{CLIP-IG} & \textcolor{blue}{Two Mickey Mouse statues} standing in front of a large room, 3 d cartoon, lee madgwick \&amp, disney pixar 3d style, pixar and disney 3d style... \cr \midrule
\textbf{PromptStealer} & \textcolor{blue}{Mickey Mouse} in a movie theater, artstation, highly detailed, sharp focus, 8k, octane render, 4k, cinematic, hd, unreal engine 5... \cr \midrule
\textbf{PH2P} & minion minion bunny dnd " amas expectations monument \cr \midrule
\textbf{VLMasExpert} & \textcolor{blue}{Two} cartoon characters resembling anthropomorphic \textcolor{blue}{mice} stand facing each other on a stage in a grand theater setting, both are wearing black outfits with large round ears and gloved hands, the background is filled with rows of empty seats in a spacious \textcolor{blue}{auditorium}... \cr \midrule
\rowcolor{lightgray}\textbf{Prometheus (Ours)} & \textcolor{blue}{Two Mickey Mice} are \textcolor{blue}{talking} on stage, \textcolor{blue}{orchestra}, \textcolor{blue}{pack auditorium}, \textcolor{blue}{Mickey Mouse statue}, \textcolor{blue}{auditorium}, \textcolor{blue}{hyper realism}. \cr
\bottomrule
\end{tabular}
 \vspace{-0.1in}
\end{table}

\para{Implementation Details.} We implement \sys using Python 3.8 with PyTorch. All experiments are conducted on a single NVIDIA GeForce RTX A6000 GPU.
The fidelity gain threshold in Equation~\ref{eq:score function} is set to 0.005. For BLIP used across all methods, we utilize the official implementation and adopt a ViT-based backbone, which is fine-tuned on the COCO dataset~\cite{lin2014microsoft}. Unless otherwise specified, the \texttt{Oracle} query limit for \sys is set to 200, and the number of captions is set to 400. Note that, since we employ a local proxy model for feedback, the money cost is 0. To tackle the randomness within the attack process and image generation, we run each experiment three times. 

\subsection{Main Result}
\label{sec: main result}
We first compare \sys with five baselines in terms of two objectives depicted in Section~\ref{sec:attack-req}: \textit{functionality} and \textit{reusability}. Additionally, we consider a practical scenario in which multiple showcases, each featuring different subjects, are provided. 

\para{Functionality.} 
Table~\ref{tab: main result} shows the main experimental results. 
In general, \sys surpasses all the baselines on both DALLEPrompt and RealPrompt, demonstrating its functionality and practicality in real-world scenarios. Specifically, it achieves the best average scores: image semantic similarity (0.897), image perception similarity (0.601), prompt semantic similarity (0.803), and ASR (56.0\%) on DALLEPrompt. Similarly, on RealPrompt, \sys leads with image semantic similarity (0.912), image perception similarity (0.625), prompt semantic similarity (0.814), and ASR (62.5\%), demonstrating its effectiveness in real-world scenarios. 

\begin{figure}[t!]
  \centering
  \includegraphics[width=\linewidth]{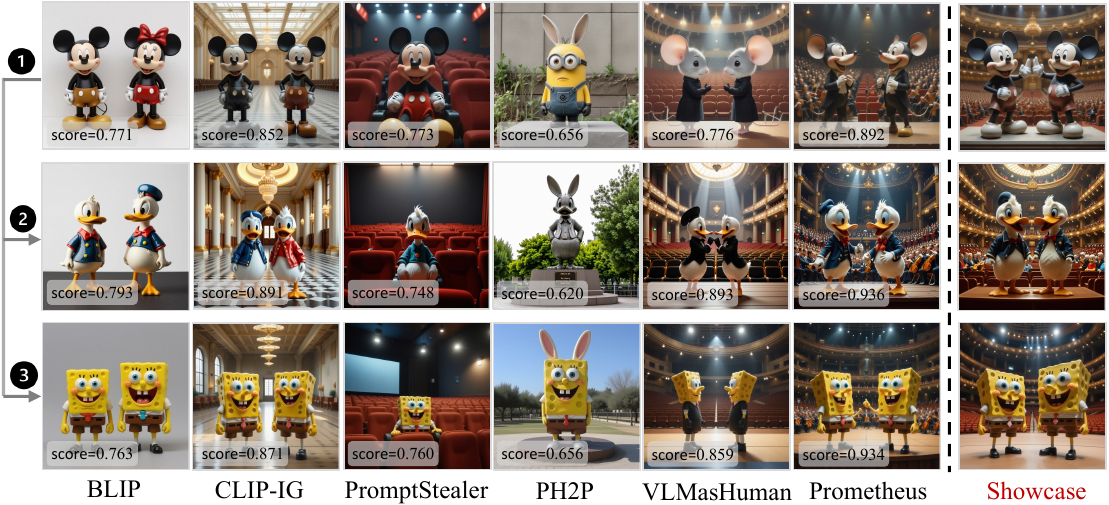} 
  \scriptsize
  \vspace{-0.1in}
  \caption{Reusability of stolen prompts with different subjects. We swap all the subject-related words (\ding{182} Mickey Mouse; \ding{183} Donald Duck; \ding{184} SpongeBob) to the target subject only. The corresponding prompts can be found in Table~\ref{tab: example}. Please zoom out for details. }
  \vspace{-0.1in}
  \label{fig: transfer-1}
\end{figure}

For image semantics and perception, \sys's improvements are mainly attributed to its prompt optimization with feedback from \textit{Proxy}, while other methods only leverage local models to predict the prompt. This feedback-based refinement helps \sys generate images that are perceptually similar to the showcase. For prompt semantics, \sys also achieves better results on all datasets and proxy models. The main reason is that we improve the quality of recovered prompts by introducing dynamic modifiers and contextual matching. The former adds subject details as well as position information, and the latter ensures the sorted modifier achieves more consistent fidelity improvement when being put into the base prompt. Therefore, \sys obtains higher prompt semantic similarity with the victim prompt. We include a more detailed analysis of the performance gap between DALLEPrompt and RealPrompt in Appendix~\ref{appendix: more analysis}.

\begin{table}[ht!]
\centering
\caption{Reusability of prompts (with different subjects) generated with \sys compared with five baselines. The evaluation dataset is the real-world dataset RealPrompt. }
\vspace{-0.1in}
\label{tab: reusability}
\resizebox*{\linewidth}{!}{
\begin{tabular}{c|cccc}
\toprule
\textbf{Method}&\textbf{${CLIP}_{img}(\uparrow)$}&\textbf{$LPIPS(\downarrow)$}&\textbf{$SBERT(\uparrow)$}&\textbf{$ASR(\uparrow)$}  \cr \midrule
BLIP&0.755&0.765&0.749&29.2\%  \\
CLIP-IG&0.785&0.725&0.779&22.9\%  \\
PromptStealer&0.775&0.756&0.809&54.2\% \\
PH2P&0.730&0.748&0.627&0\% \\
VLMasExpert&0.767&0.715&0.801&47.9\% \\
\rowcolor{lightgray}Prometheus (Ours)&\textbf{0.832}&\textbf{0.681}&\textbf{0.835}&\textbf{70.8\%} \cr \bottomrule
\end{tabular}
}
\end{table}

Figure~\ref{fig: visualization-1} illustrates eight examples of the attack effect achieved by \sys compared to baseline methods. Additionally, Table~\ref{tab: example} presents an example of the stolen prompts. Overall, compared to the baselines, \sys generates prompts and images with higher fidelity to the ground truths and showcases in terms of image semantics, visual perception, and prompt semantics. 
Take Figure~\ref{fig: visualization-1} case VI as an example. The baseline BLIP struggles to recognize the style (\textit{e.g.}, ``hyperrealistic'') and fails to capture the main content of the image (\textit{e.g.}, ``in a concert hall''). Meanwhile, CLIP-IG tends to match irrelevant modifiers to the showcase and introduces excessive redundancy. For instance, it ranks ``3D cartoon'' as having the highest correlation to the showcase (VI) and includes many unrelated ``Disney'' modifiers. These mismatches result in semantically incorrect image content and lower perceptual quality. This behavior can be attributed to CLIP-IG treating modifiers independently, leading to poorly contextualized predictions. 
PromptStealer exhibits two notable shortcomings. First, it demonstrates lower fidelity when predicting the subject (\textit{e.g.}, Mickey Mouse in case II and The Simpsons in case VII). Second, as shown in Table~\ref{tab: example}, PromptStealer often predicts generic modifiers that are frequent in its training data~\cite{PromptStealing}, rather than contextually appropriate ones. A likely explanation is that PromptStealer’s prediction models are trained on out-of-domain data, such as that collected from Stable Diffusion, leading to overfitting. 
As for PH2P~\cite{mahajan2024prompting}, this method fails to capture style modifiers and often even the subject. This highlights a fundamental limitation of using gradient-based prompt inversion for prompt stealing, indicating the need for further refinement to enable adaptation.
Another baseline, VLMasExpert, performs relatively well but frequently omits key characters in the showcase. This limitation is likely due to alignment constraints during the model training and the out-of-distribution problem. 
We provide a more detailed analysis about the impact of alignment on its stealing performance in Appendix~\ref{appendix: more results}. 
In contrast, \sys captures dynamic and contextually relevant modifiers, such as ``auditorium'' and ``orchestra'', which align with elements present in the showcase. As we can see, in the base prompt, \sys also ignores the ``concert hall'' element, but the dynamic modifier fixes this issue. Furthermore, \sys accurately identifies the modifier ``hyper realism'', demonstrating its outstanding prediction ability.

\begin{table}[t!]
\centering
\caption{Performance under multiple showcase scenarios. The evaluation dataset is the real-world dataset RealPrompt.}
\vspace{-0.1in}
\label{tab: multiple}
\resizebox*{\linewidth}{!}{
\begin{tabular}{c|cccc}
\toprule
\textbf{Method}&\textbf{${CLIP}_{img}(\uparrow)$}&\textbf{$LPIPS(\downarrow)$}&\textbf{$SBERT(\uparrow)$}&\textbf{$ASR(\uparrow)$}  \cr \midrule
BLIP&0.778&0.760&0.760&27.1\%  \\
CLIP-IG&0.808&0.723&0.805&72.9\%  \\
PromptStealer&0.796&0.742&0.809&75.0\% \\
PH2P&0.743&0.737&0.651&0\% \\
VLMasExpert&0.803&0.715&0.801&58.3\% \\
\rowcolor{lightgray}Prometheus (Ours)&\textbf{0.834}&\textbf{0.684}&\textbf{0.840}&\textbf{79.2\%} \cr \bottomrule
\end{tabular}
}
\vspace{-0.1in}
\end{table}

\para{Reusability on Different Subjects.} Table~\ref{tab: reusability} presents the comparison of reusability between \sys and baseline methods. Reusability is a crucial property, as it enables users to modify the subject of a prompt while maintaining similar styles in the generated images. To evaluate this, we use the RealPrompt dataset, which comprises 8 style sets (three subjects per style) to simulate real-world scenarios. For each stolen prompt, an LLM is employed to replace the subject, producing a new prompt with a different subject. This process is achieved using the following instruction with GPT-4o: ``Please replace the central subject in the sentence {prompt} with {subject} while keeping the rest of the content unchanged''. We validate the reusability of \sys and other baselines from both qualitative and quantitative angles. We present the experimental results in Table~\ref{tab: reusability} and visualization in Figure~\ref{fig: transfer-1}. 

\begin{figure}[t!]
    \centering
    \subfigure[FLUX]{%
        \includegraphics[width=0.31\linewidth]{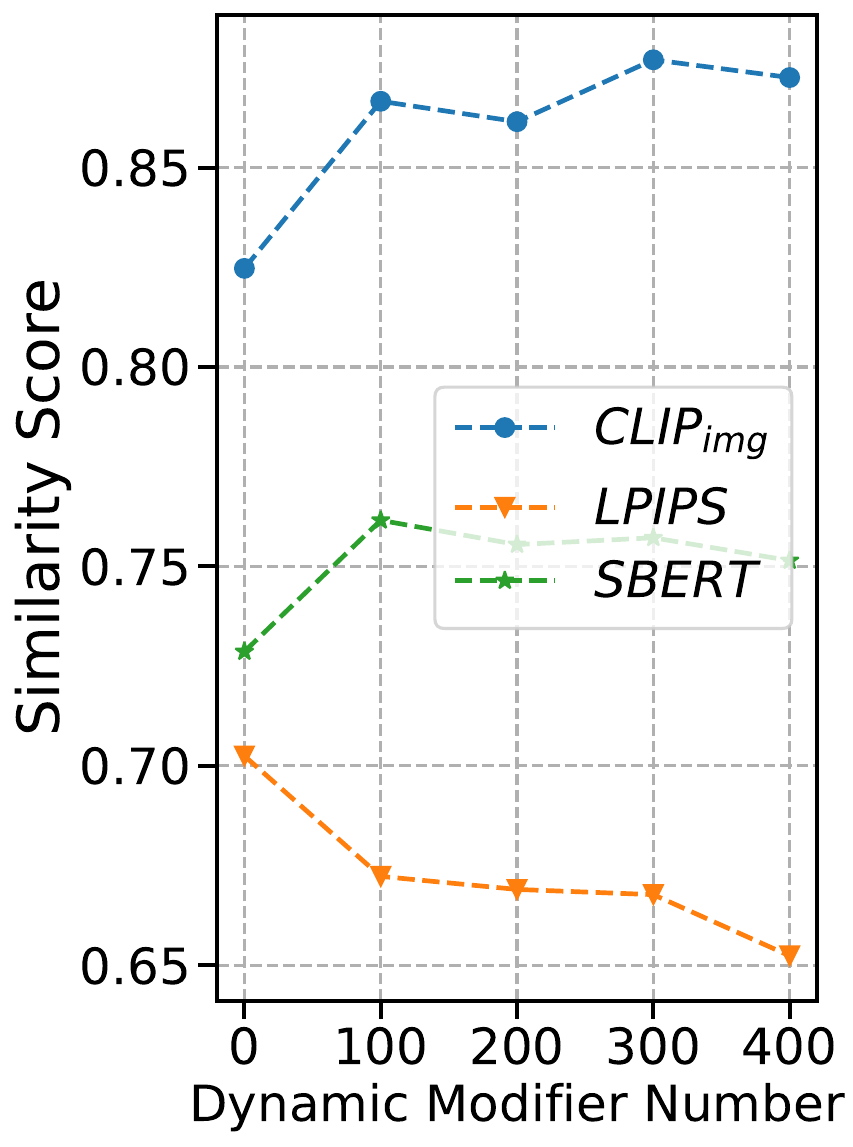}
        \label{fig: abla-dynamic-1}
    }
    \subfigure[ShuttleDiffusion]{%
        \includegraphics[width=0.31\linewidth]{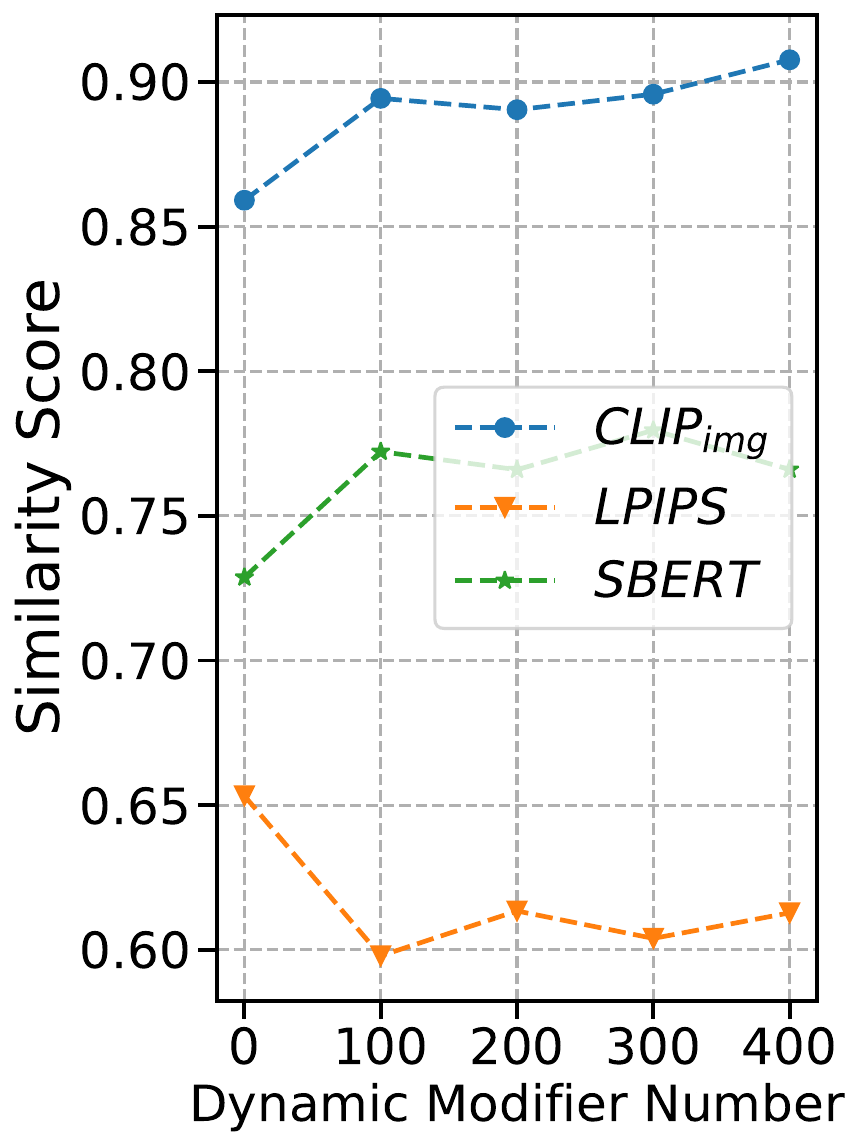}
        \label{fig: abla-dynamic-2}
    }
    \subfigure[SD-3.5]{%
        \includegraphics[width=0.31\linewidth]{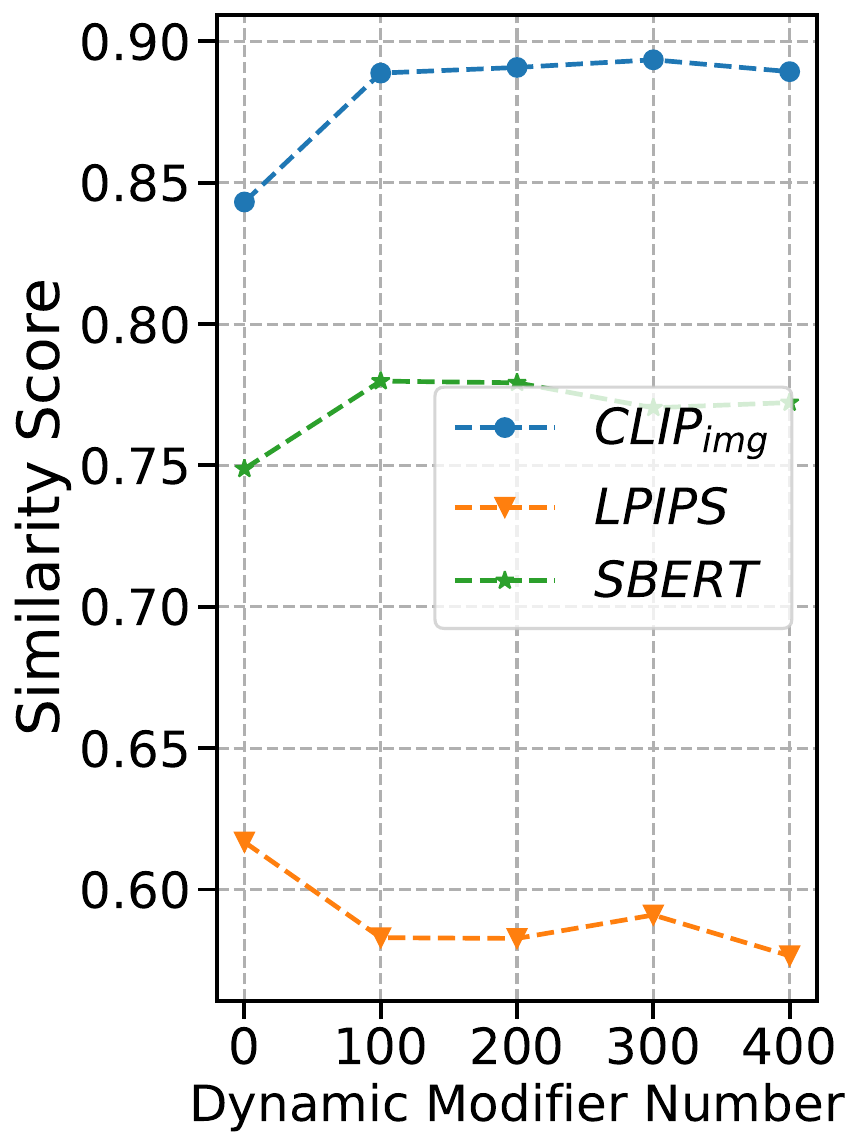}
        \label{fig: abla-dynamic-3}
    }
    \vspace{-0.1in}
    \caption{Impact of the number of extracted dynamic modifiers. The dataset is DALLEPrompt (${CLIP}_{img}:\uparrow$, $LPIPS:\downarrow$, $SBERT:\uparrow$). }
    \vspace{-0.1in}
    \label{fig: ablation-dynamic-a}
\end{figure}

As shown in Table~\ref{tab: reusability}, \sys outperforms all baselines across metrics, including image semantics, image perception, prompt semantics, and ASR. This demonstrates that our method extracts the most reusable prompts while minimizing overfitting to specific showcases. 
We illustrate this in Figure~\ref{fig: transfer-1}, where the subject is replaced from ``Mickey Mouse'' to ``Donald Duck'' and ``SpongeBob''. As shown, \sys maintains a consistent style that aligns closely with the ground truth outcomes, whereas the baselines exhibit several shortcomings. Specifically, BLIP, CLIP-IG, and PH2P fail to capture the ``concert hall'' element, resulting in images with incorrect semantics. PromptStealer partially captures the subject but introduces errors in quantity and mismatches static modifiers. In contrast, \sys achieves superior image semantics and perceptual similarity, demonstrating its effectiveness in generating reusable prompts.
\begin{mdframed}[nobreak=true]
\textit{Take-away}: \sys outperforms the existing prompt stealing attack~\cite{PromptStealing} and potential adaptations in both \textit{functionality} and \textit{reusability} with a large margin, indicating its high effectiveness. 
\end{mdframed}

\para{Multiple Showcases.} We consider a practical scenario where multiple showcases are provided under a single theme. In this case, the attacker extracts a prompt for each showcase, resulting in a set of prompts. These prompts are then merged into a distilled prompt. Here, we consider using a large language model (\textit{i.e.,} ChatGPT-4o) to summarize the stolen prompts with the instruction ``Please summarize these sentences into one sentence by rephrasing them. Replace $\left\{subjects\right\}$ or related words with $\left\{target \ subject\right\}$''. The experimental results are presented in Table~\ref{tab: multiple}, leading to two key observations. First, while \sys continues to outperform the baseline methods, the margin of improvement is less pronounced. A potential explanation is that the baseline methods compensate for inadequacies in single stolen prompts by leveraging the collective information from the entire prompt set, thereby reducing \sys's relative advantage. Second, compared to the results in Table~\ref{tab: main result}, the prompt semantic fidelity of \sys improves (from 0.814 to 0.840) after merging all the stolen prompts. This is likely because the summarization process mitigates overfitting to individual showcases, resulting in a stolen prompt that is better aligned with the ground truth. 
\begin{mdframed}[nobreak=true]
\textit{Take-away}: When given multiple showcases on the same theme, \sys outperforms the baselines and achieves better performance than in single-showcase scenarios.  
\end{mdframed}

\begin{figure}[t!]
    \centering
    \subfigure[FLUX]{%
        \includegraphics[width=0.31\linewidth]{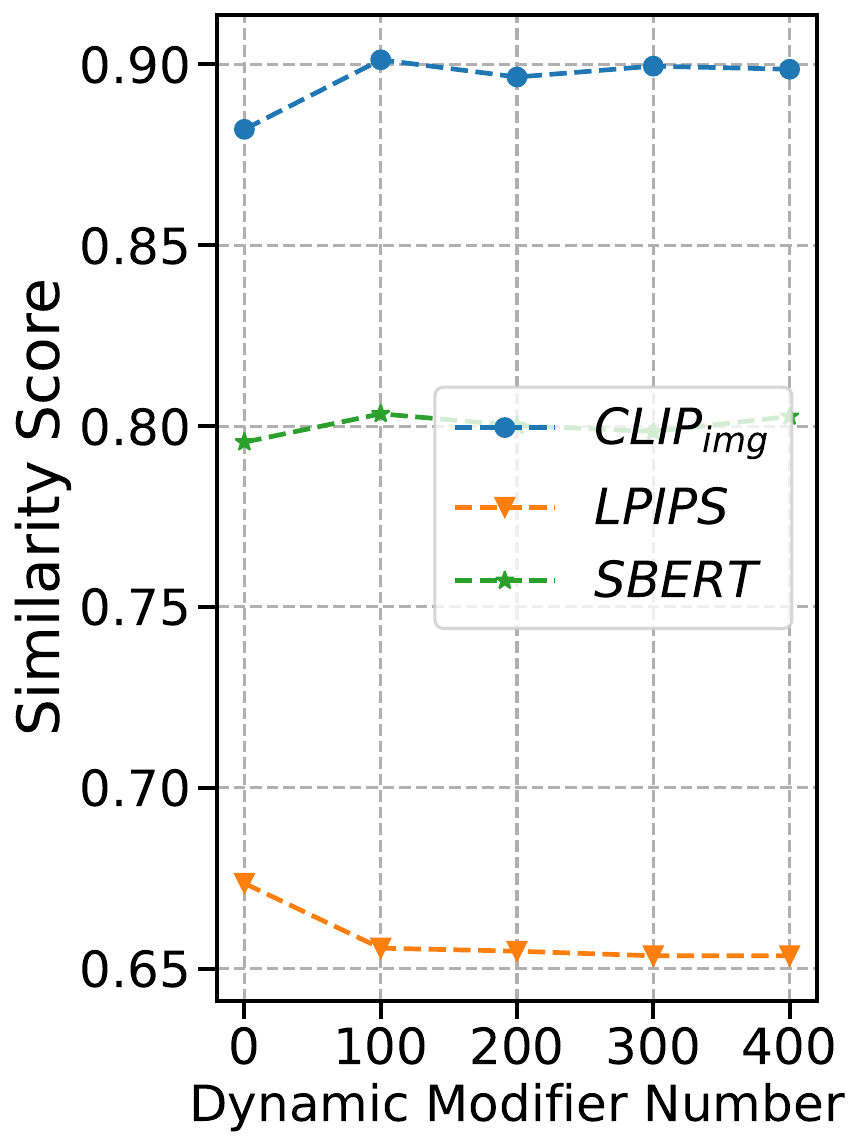}
        \label{fig: abla-dynamic-4}
    }
    \subfigure[ShuttleDiffusion]{%
        \includegraphics[width=0.31\linewidth]{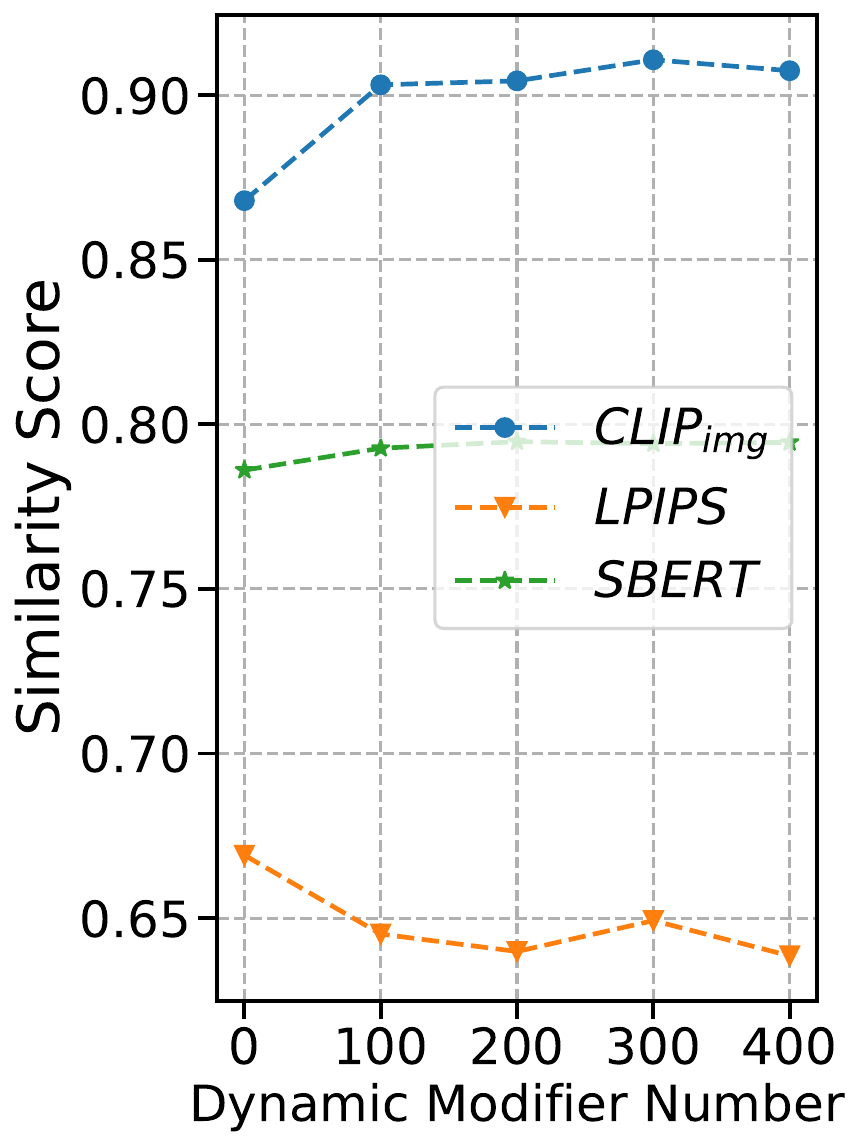}
        \label{fig: abla-dynamic-5}
    }
    \subfigure[SD-3.5]{%
        \includegraphics[width=0.31\linewidth]{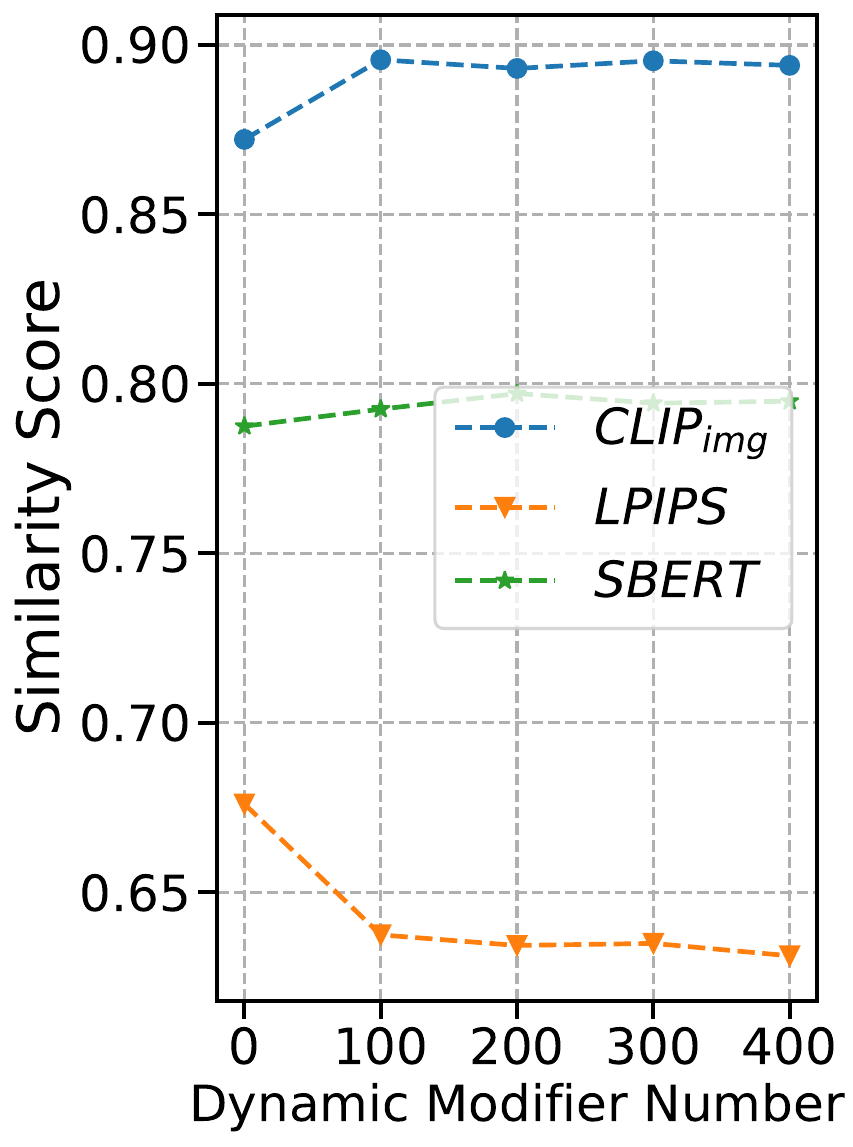}
        \label{fig: abla-dynamic-6}
    }
    \vspace{-0.1in}
    \caption{Impact of the number of extracted dynamic modifiers. The evaluation dataset is RealPrompt (${CLIP}_{img}:\uparrow$, $LPIPS:\downarrow$, $SBERT:\uparrow$). }
    \vspace{-0.1in}
    \label{fig: ablation-dynamic-b}
\end{figure}

\subsection{Ablation Study}
\label{sec: ablation study}

\begin{figure*}[ht!]
    \centering
    \subfigure[FLUX]{%
        \includegraphics[width=0.27\linewidth]{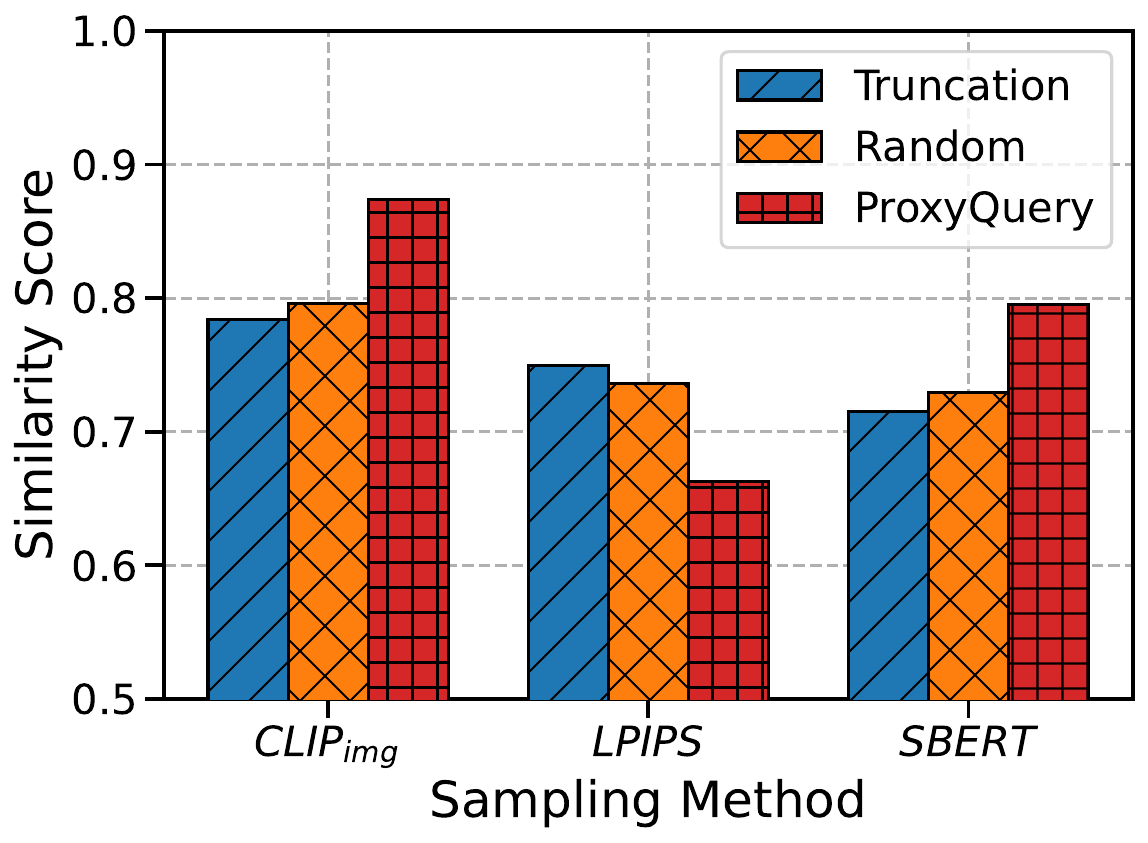}
        \label{fig: abla-proxy-1}
    }
    \subfigure[ShuttleDiffusion]{%
        \includegraphics[width=0.27\linewidth]{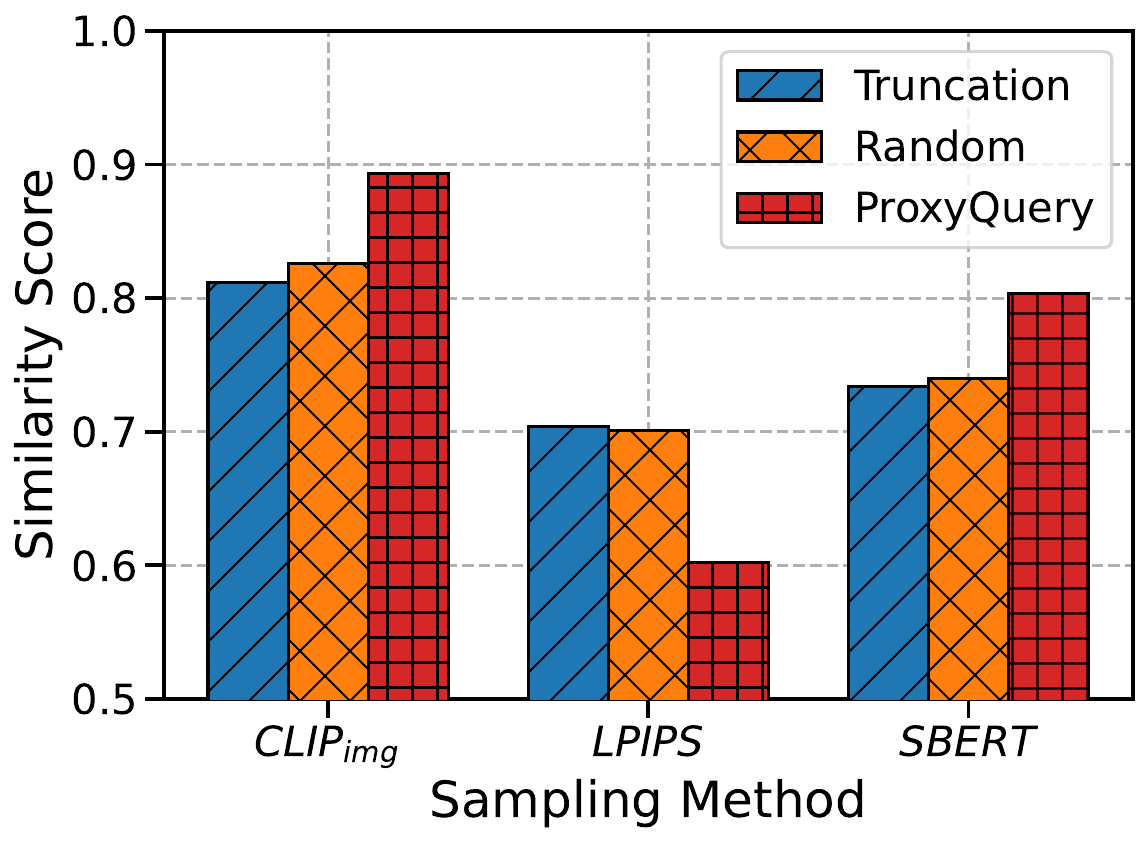}
        \label{fig: abla-proxy-2}
    }
    \subfigure[SD-3.5]{%
        \includegraphics[width=0.27\linewidth]{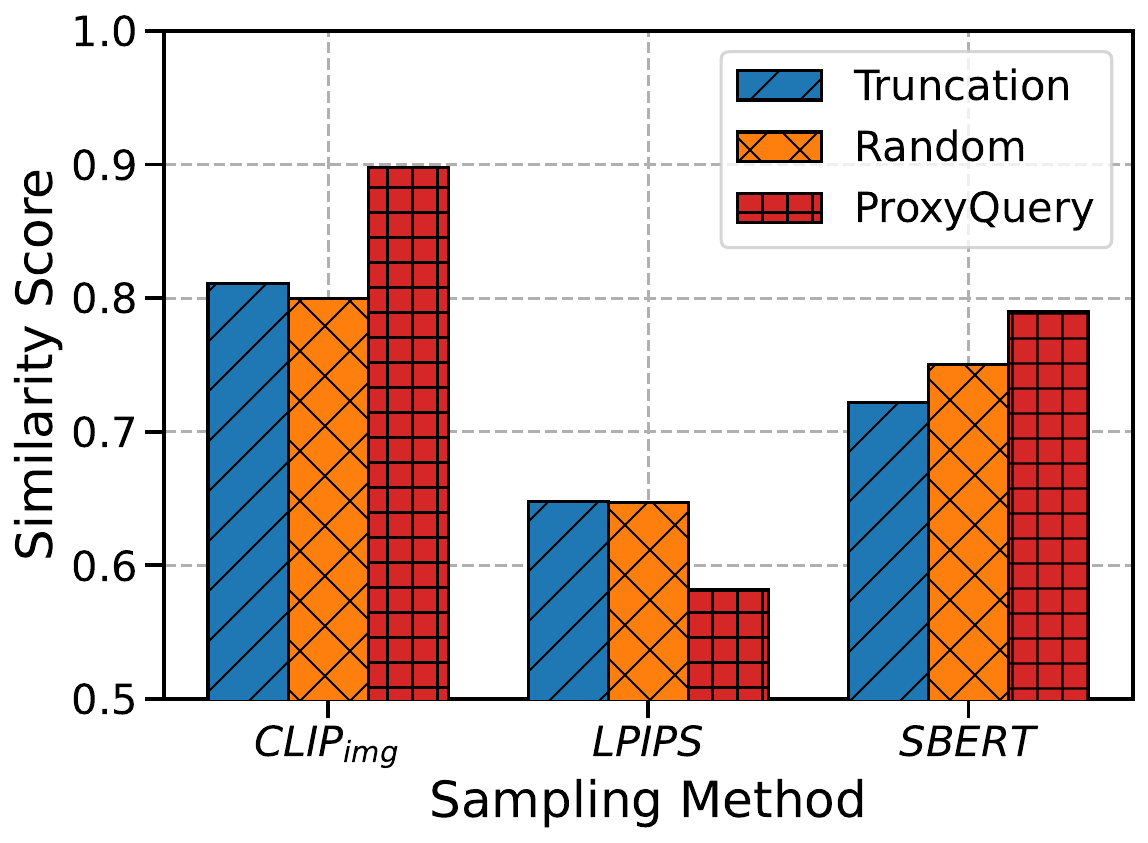}
        \label{fig: abla-proxy-3}
    }
    \vspace{-0.1in}
    \caption{Impact of \texttt{Proxy} query. The evaluation dataset is DALLEPrompt (${CLIP}_{img}:\uparrow$, $LPIPS:\downarrow$, $SBERT:\uparrow$). }
    \vspace{-0.1in}
    \label{fig: ablation-proxy-a}
\end{figure*}

\begin{figure*}[ht!]
    \centering
    \subfigure[FLUX]{%
        \includegraphics[width=0.27\linewidth]{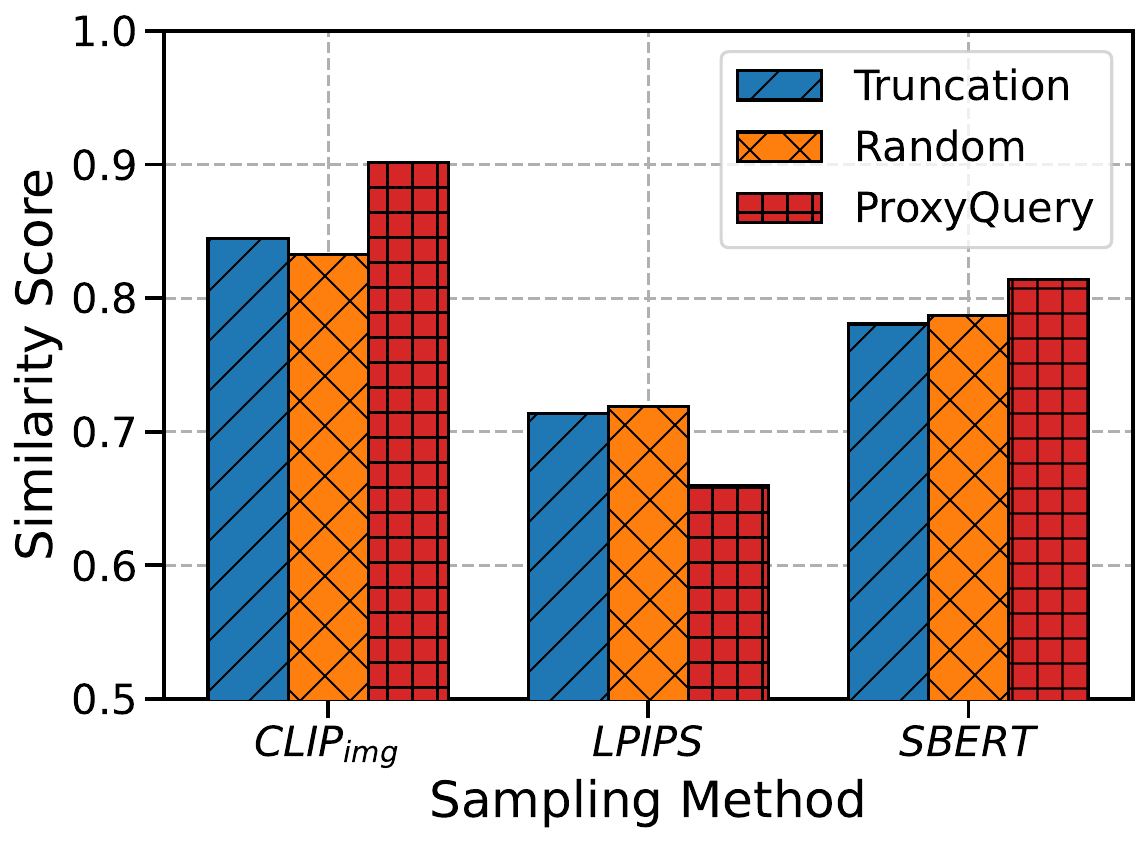}
        \label{fig: abla-proxy-4}
    }
    \subfigure[ShuttleDiffusion]{%
        \includegraphics[width=0.27\linewidth]{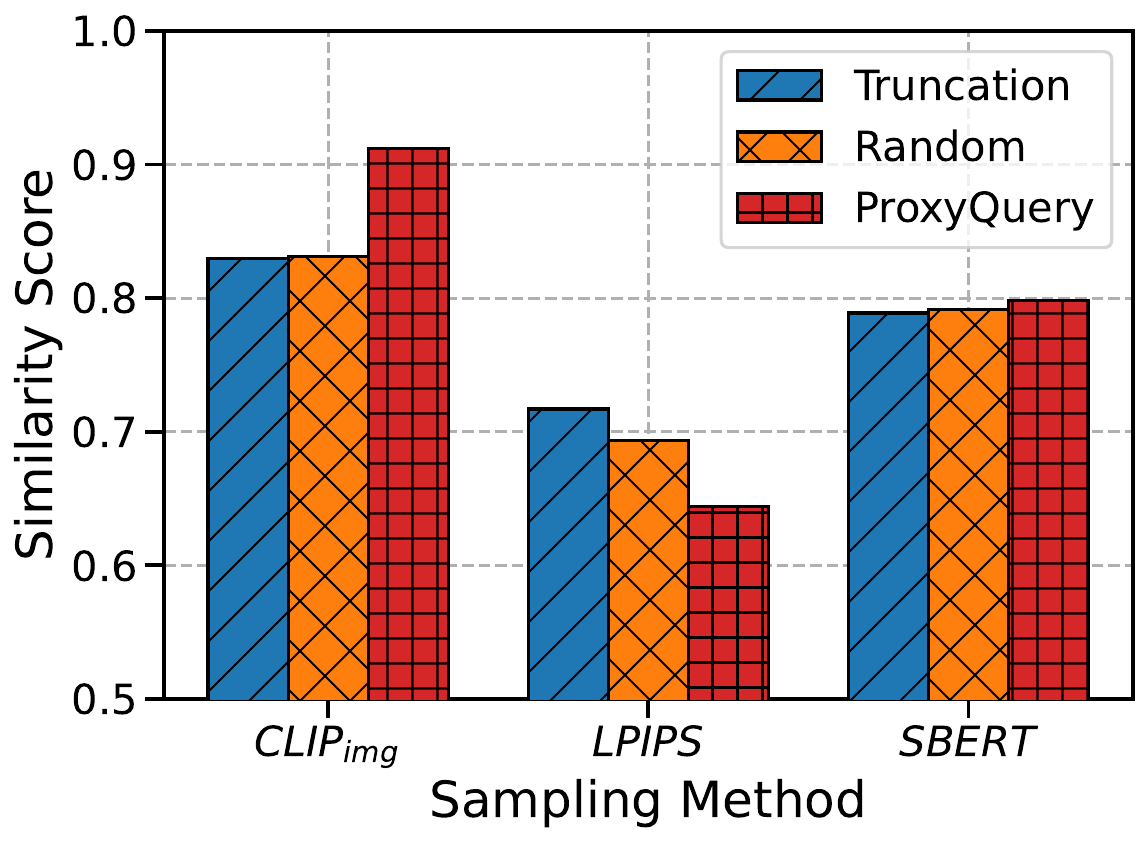}
        \label{fig: abla-proxy-5}
    }
    \subfigure[SD-3.5]{%
        \includegraphics[width=0.27\linewidth]{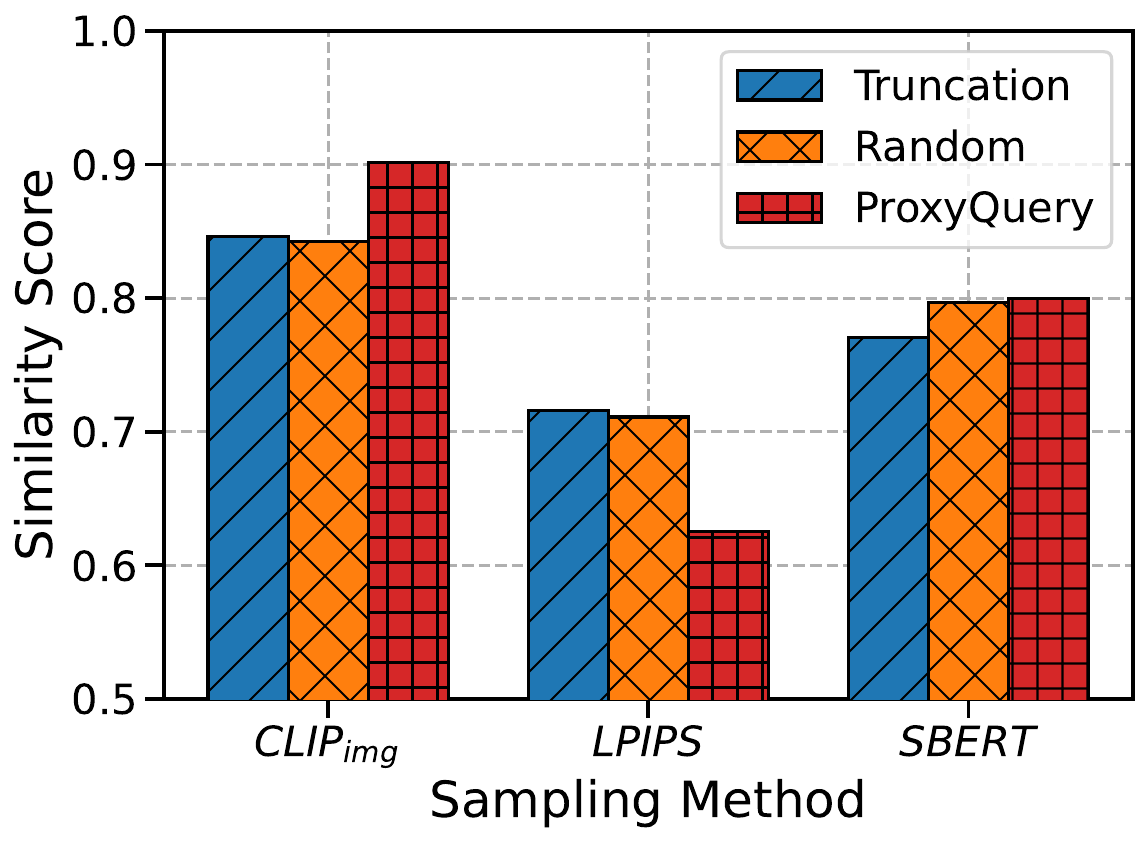}
        \label{fig: abla-proxy-6}
    }
    \vspace{-0.1in}
    \caption{Impact of \texttt{Proxy} query. The evaluation dataset is RealPrompt (${CLIP}_{img}:\uparrow$, $LPIPS:\downarrow$, $SBERT:\uparrow$).}
    \vspace{-0.1in}
    \label{fig: ablation-proxy-b}
\end{figure*}

\para{Impact of Dynamic Modifiers.} Figure~\ref{fig: ablation-dynamic-a} and \ref{fig: ablation-dynamic-b} show the impact of the number of captions that are used to generate dynamic modifiers, ranging from 0 to 400, on \sys's performance with three metrics. We have two observations. First, including dynamic modifiers boosts the stealing performance. As shown in all cases, when adopting dynamic modifiers (quantity = 100), the similarities in the three metrics outperform the method that excludes dynamic modifiers (quantity = 0). The reason is that our proposed dynamic modifiers enhance the prompt's detail and positional information, providing a better description of the showcase. Second, \sys shows a stable performance trend as the modifier quantity increases from 100 to 400. The potential reason is that \sys extracts high-quality dynamic modifiers from the captions of the showcase, requiring only a limited number of captions. This reveals the high efficiency of \sys. 

\begin{table}[ht!]
\caption{Impact of specific dynamic modifiers. Here we consider removing subject detail or position information to study which plays a more important role in improving overall performance. }
\label{tab: ablation-specific-modifier}
\centering
\begin{tabular}{cc|ccc}
\toprule
\textbf{Detail}&\textbf{Position}&${CLIP}_{img}(\uparrow)$&$LPIPS(\downarrow)$&$SBERT(\uparrow)$  \cr \midrule
\usym{2718}&\usym{2718}&0.882&0.674&0.796  \\
\usym{2714}&\usym{2718}&0.897&0.659&0.810 \\ 
\usym{2718}&\usym{2714}&0.891&0.665&0.807 \\
\usym{2714}&\usym{2714}&\textbf{0.901}&\textbf{0.653}&\textbf{0.814}  \\ \bottomrule
\end{tabular}
\end{table}

\para{Impact of Specific Dynamic Modifiers.} We further validate the effectiveness of specific components of dynamic modifiers, \ie, \textit{subject detail} and \textit{position information}. Here, we consider \sys with only one kind of dynamic modifiers by excluding the other kind during modifier collection. As shown in Table~\ref{tab: ablation-specific-modifier}, we have two observations. First, including one or both types of dynamic modifiers improves performance. Subject detail and position information both have gains in image appearance and prompt fidelity, and the gains are even greater when they are both included. Second, subject detail provides more gain than position details. The position information includes position details and azimuth information, where the former is covered by the subject detail to a certain degree. 

\begin{table}[t!]
\caption{Impact of matching method. We replace the contextual matching in \sys with the vanilla method. }
\vspace{-0.1in}
\label{tab: ablation-in-context}
\resizebox*{\linewidth}{!}{
\begin{tabular}{cc|ccc}
\toprule
\textbf{\textit{Proxy}}&\textbf{Methods}&${CLIP}_{img}(\uparrow)$&$LPIPS(\downarrow)$&$SBERT(\uparrow)$  \cr \midrule
\multirow{2}{*}{FLUX} &Vanilla&0.895&0.659&0.789  \\
&Ours&\textbf{0.901}&\textbf{0.653}&\textbf{0.814}  \\ \midrule
\multirow{2}{*}{ShuttleDiffusion} &Vanilla&0.902&0.655&0.796  \\
&Ours&\textbf{0.912}&\textbf{0.644}&\textbf{0.798}  \\ \midrule
\multirow{2}{*}{SD-3.5} &Vanilla&0.893&0.642&0.793  \\
&Ours&\textbf{0.901}&\textbf{0.625}&\textbf{0.799}  \\ \bottomrule

\end{tabular}
}
\vspace{-0.1in}
\end{table}

\para{Impact of Contextual Matching.} We replace our proposed contextual matching method in \sys with a vanilla approach that directly calculates cosine similarity between the modifiers and showcases~\cite{Interrogator}. Table~\ref{tab: ablation-in-context} presents the results across three metrics. As the matching method transitions from vanilla to our approach, the attack performance improves. Notably, our contextual matching outperforms the vanilla method across all proxy models. A key highlight is the significant increase in prompt semantic similarity for prompts stolen with FLUX, rising from 0.789 to 0.814. The core reason is that our contextual matching considers modifiers' contextual semantics so that it improves the prompt similarity. 

\begin{figure}[ht!]
    \centering
    \subfigure[FLUX]{%
        \includegraphics[width=0.31\linewidth]{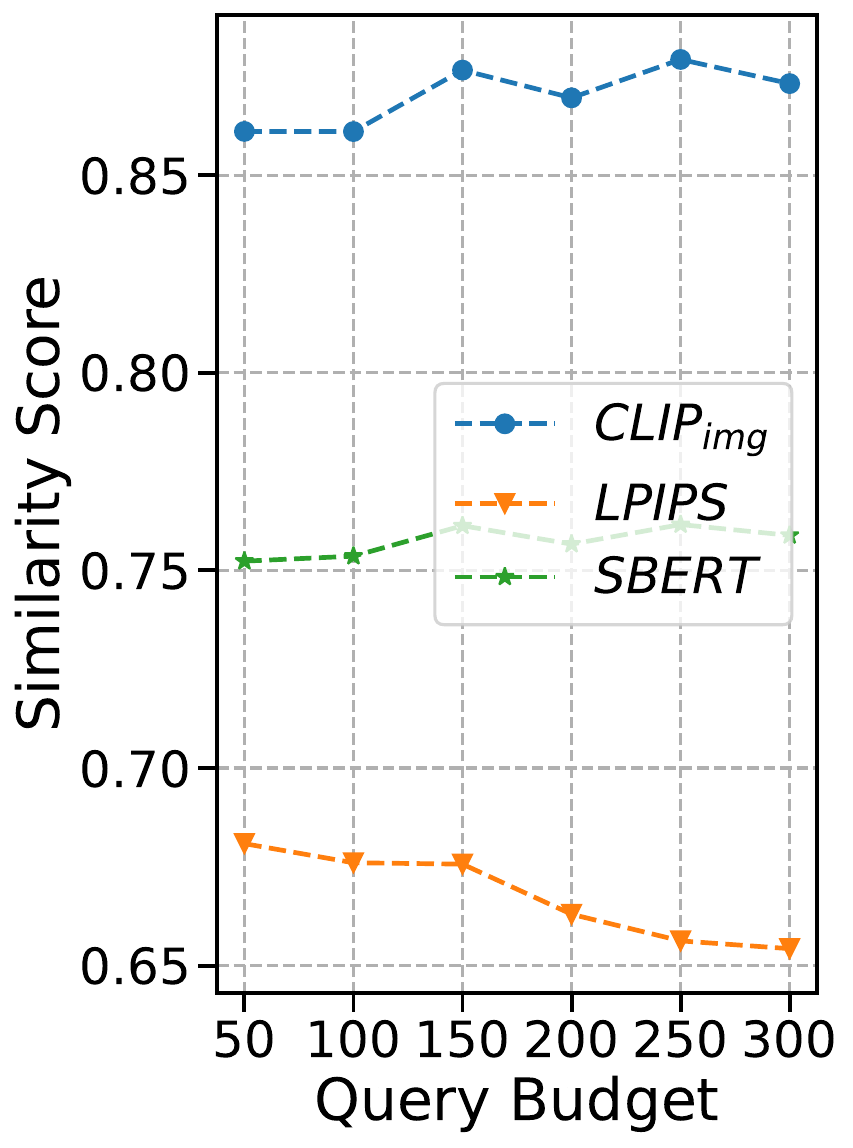}
        \label{fig: ablation-budget-1}
    }
    \subfigure[ShuttleDiffusion]{%
        \includegraphics[width=0.31\linewidth]{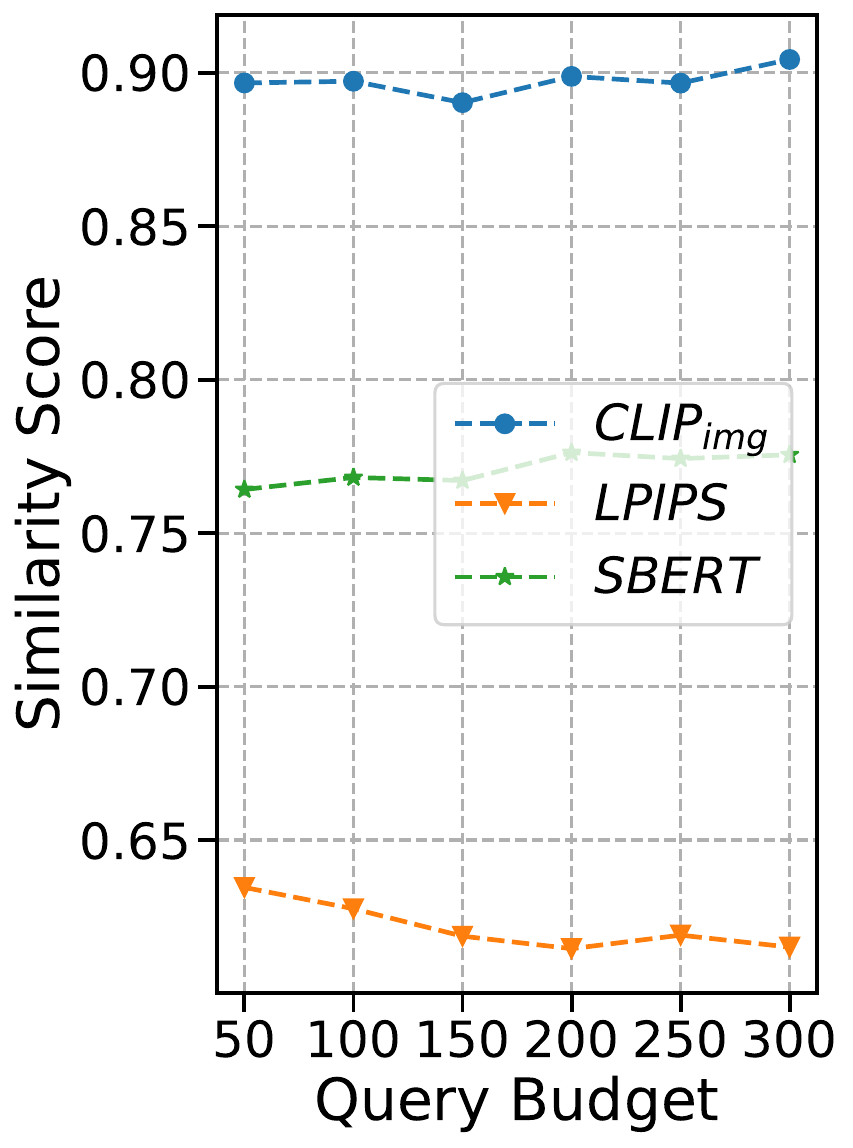}
        \label{fig: ablation-budget-2}
    }
    \subfigure[SD-3.5]{%
        \includegraphics[width=0.31\linewidth]{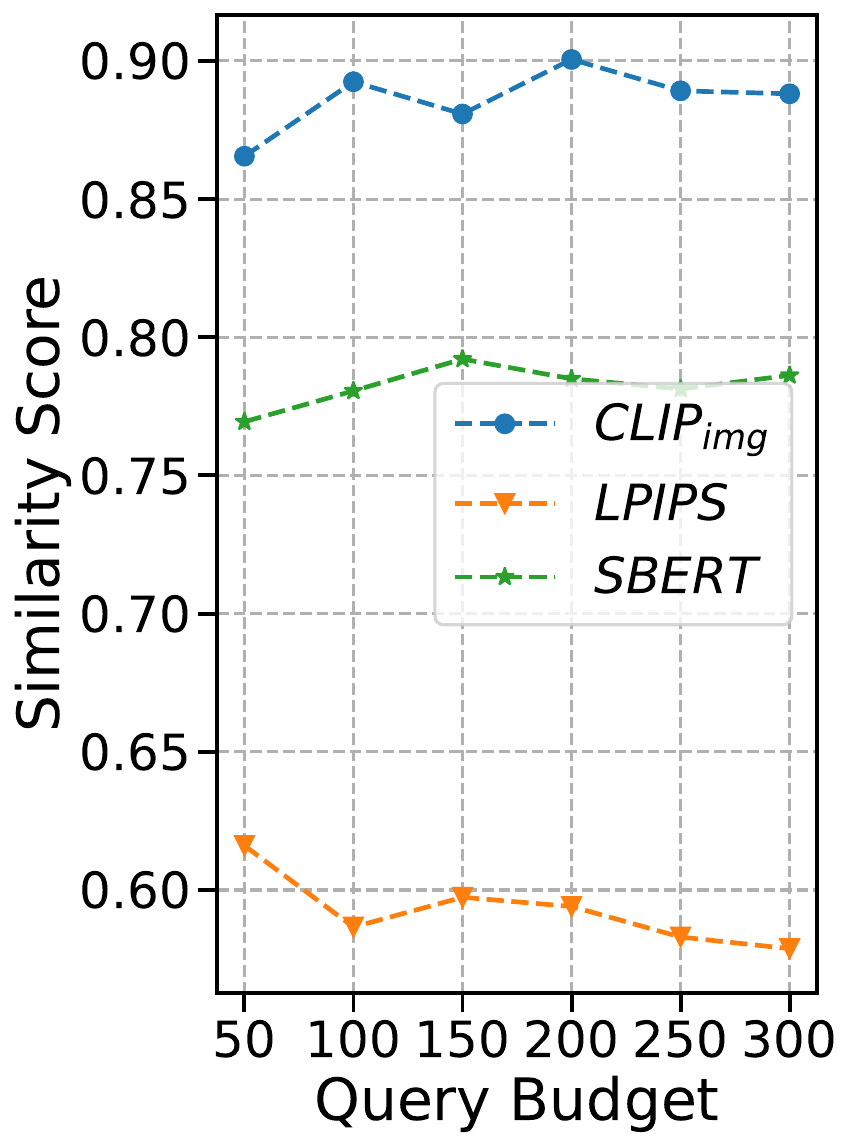}
        \label{fig: ablation-budget-3}
    }
    \vspace{-0.1in}
    \caption{Impact of number of $\texttt{Proxy}$ query. The dataset is DALLEPrompt (${CLIP}_{img}:\uparrow$, $LPIPS:\downarrow$, $SBERT:\uparrow$). }
    \vspace{-0.1in}
    \label{fig: ablation-budget-a}
\end{figure}

\para{Impact of \texttt{Proxy} Feedback.} We now illustrate how the \texttt{Proxy} feedback influences the performance of \sys. We consider two baselines to replace our greedy search, \textit{i.e.,} direct splicing and random sampling. Direct splicing implies that the attacker directly splices the top-$k$ modifiers with the highest correlation to the subject, while random sampling chooses $k$ modifiers at random. We set $k$ to 20 for these two baselines and the sampling pool of random sampling to the best 200 modifiers, in line with the query budget of \sys. Figures~\ref{fig: ablation-proxy-a} and \ref{fig: ablation-proxy-b} show the comparison results on three metrics. As all the modifiers are refined with our local contextual matching, which is based on modifier-showcase cosine similarity, all the methods achieve a rather good performance. This can be reflected in Figure~\ref{fig: ablation-budget-3}. Beyond that, we observe our proposed method significantly improves the image and prompt similarity, as shown in Figures~\ref{fig: ablation-budget-1} and \ref{fig: ablation-budget-2}. The reason is twofold. First, although our local ranking provides a highly correlated match (a correlation coefficient of 0.65), it is insufficient to directly adopt the top modifiers as the final choice, as there is still a non-negligible gap. More detailed analysis can be found in Section~\ref{sec:greed-query}. 
Second, \sys remedies this issue by leveraging the feedback from \textit{Proxy}. This helps keep the truly relevant modifiers while dropping the irrelevant ones, and thus improves the appearance alignment and prompt fidelity.

\begin{figure}[t!]
    \centering
    \subfigure[FLUX]{%
        \includegraphics[width=0.31\linewidth]{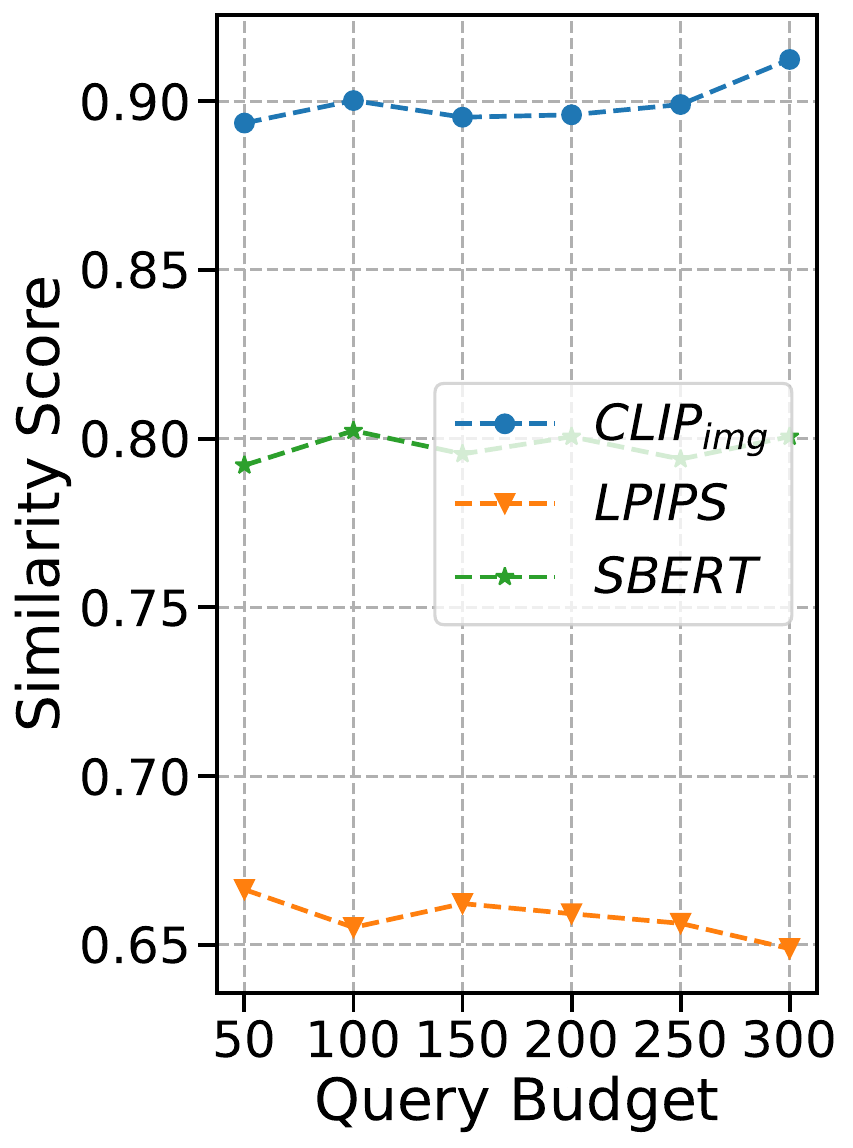}
        \label{fig: ablation-budget-4}
    }
    \subfigure[ShuttleDiffusion]{%
        \includegraphics[width=0.31\linewidth]{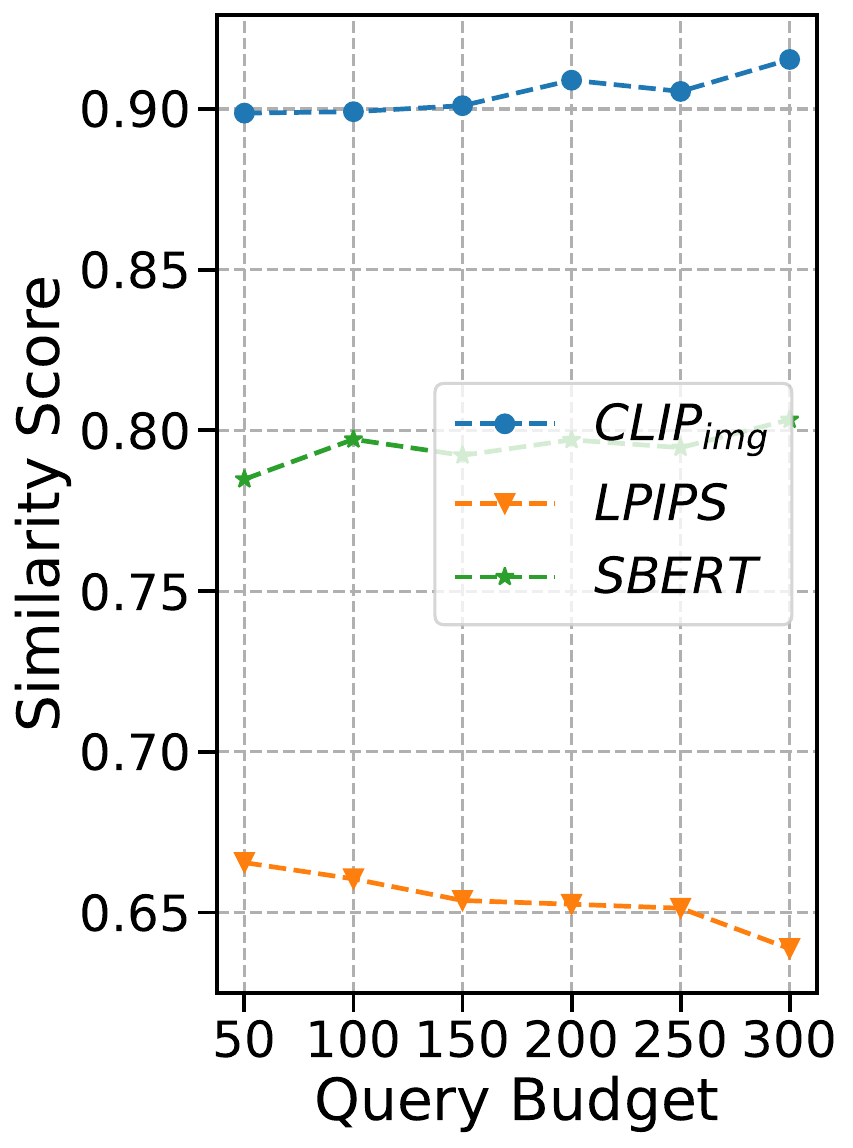}
        \label{fig: ablation-budget-5}
    }
    \subfigure[SD-3.5]{%
        \includegraphics[width=0.31\linewidth]{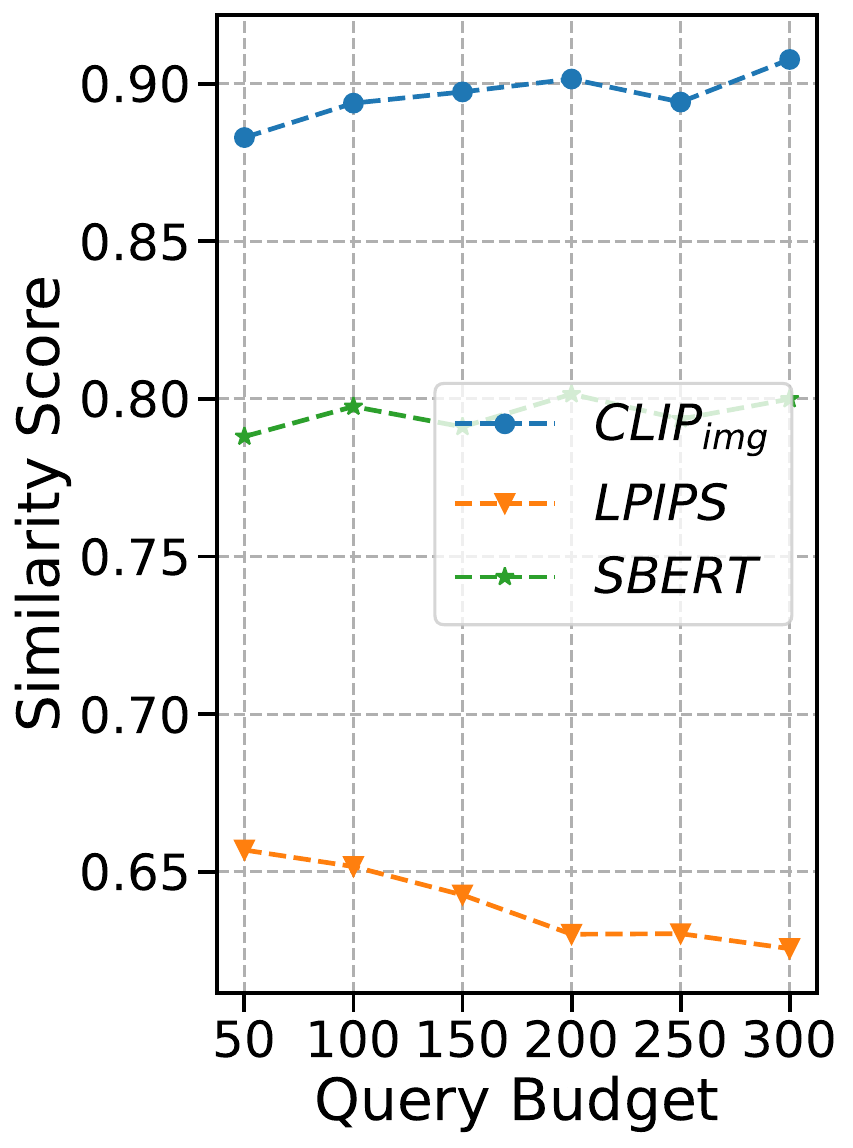}
        \label{fig: ablation-budget-6}
    }
    \vspace{-0.1in}
    \caption{Impact of the number of $\texttt{Proxy}$ query. The evaluation dataset is RealPrompt (${CLIP}_{img}:\uparrow$, $LPIPS:\downarrow$, $SBERT:\uparrow$).}
    \label{fig: ablation-budget-b}
\end{figure}

\para{Impact of Query Budget.} Figures~\ref{fig: ablation-budget-a} and \ref{fig: ablation-budget-b} illustrate the performance of \sys across all three metrics as the query budget increases from 50 to 300. We observe that when the query budget rises from 50 to 100, \sys demonstrates significant improvements in image and prompt similarities. However, the performance stabilizes as the budget increases from 100 to 300. This indicates the efficiency of \sys, as it requires relatively few queries to achieve satisfactory stealing results. This also validates the effectiveness of our local contextual matching, whose coarse-grained filtering restricts an exact range, which further facilitates the fine-grained filtering in \texttt{Proxy} query. 

\begin{mdframed}[nobreak=true]
\textit{Take-away}: \sys is efficient, requiring few captions and online \texttt{Proxy} queries. The designed modules in \sys help improve the effectiveness. 
\end{mdframed}

\subsection{User Study}
\label{sec: user study}

We conducted a user study to gain deeper insights into users' preferences regarding the recovered prompts. The question asked of the interviewees is ``How similar do you think image A is to image B?''. To mitigate potential bias, we did not inform them of our task or what these images represented. The preference is divided into five levels, which are specified in the Appendix. We conducted our investigation anonymously, and finally, we gathered 103 valid answer sheets. The detailed user study setting, IRB exemption, and our effort to protect the volunteers are depicted in Appendix~\ref{appendix: user study}. Figure~\ref{fig: user study} shows the user study results. We can tell that our \sys achieves the highest similarity according to respondents' preference, indicating its practicality in the real world. 
\begin{mdframed}[nobreak=true]
\textit{Take-away}: With sufficient ethical consideration and under fair settings, \sys achieves the highest alignment with human perceptions. 
\end{mdframed}

\begin{figure}[t!]
  \centering
  \includegraphics[width=0.9\linewidth]{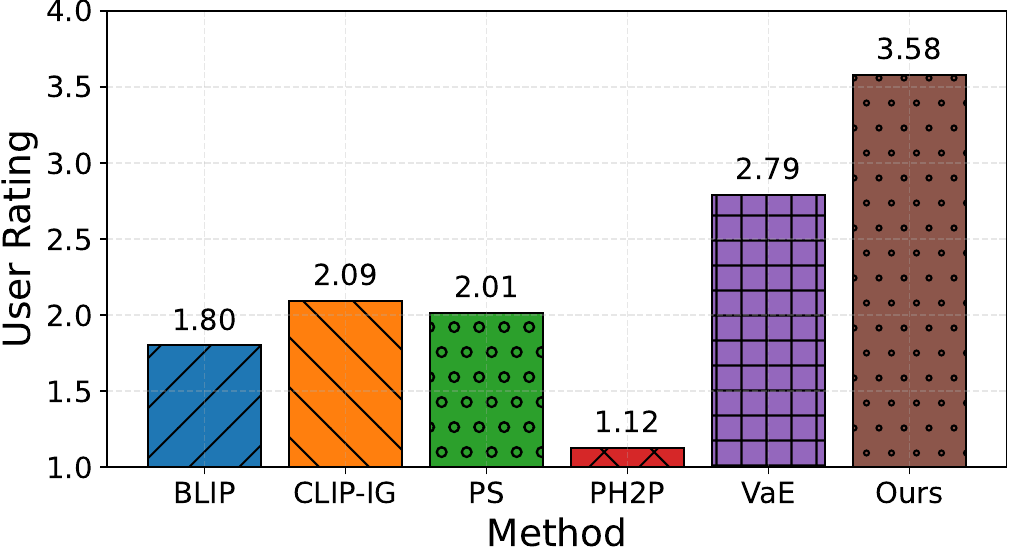} 
  \caption{Rating of user perception. PS refers to PromptStealer~\cite{PromptStealing}, and VaE refers to VLMasExpert. }
  \label{fig: user study}
  \vspace{-0.1in}
\end{figure}
\section{Potential Defenses}

\subsection{Conventional Defenses}

We first consider some existing defenses as below. 
(1) \textit{Random noise}. We generate a normally distributed noise with the same threshold of $\epsilon$ to ensure a similar visual disturbance. 
(2) \textit{Puzzle effect}. This defense covers the showcase with the puzzle effect, which is applied in practice~\cite{puzzleeffect}. 
(3) \textit{Text watermark}. This is also one defense method used in reality~\cite{textwatermark}. It covers the image with repetitive text indicating the ownership. 
We present the visual effect of these defenses in Figure~\ref{fig: defenses}. 
We do not consider PromptShield proposed in PromptStealer~\cite{PromptStealing}, as it assumes that the attacker uses a modifier prediction model to steal the prompt and the defender has white-box access to the exact modifier prediction model, which does not apply to our attack and is beyond our threat model. Instead, we have a thorough analysis in~\ref{appendix: defense}. 
We also exclude methods for prompt copyright protection (\textit{e.g.,} PromptCARE~\cite{promptcare}), as they can only verify the ownership of prompts, but cannot prevent privacy leakage. Another method, \textit{Glaze}~\cite{glaze}, is a perturbation-based method to prevent unauthorized usage of images. It optimizes an invisible perturbation to mislead the showcase feature to a \textit{target} style. Unfortunately, the code for \textit{Glaze} is not open-sourced, so we do not include it in our evaluation.

Table~\ref{tab: potential defense-1} shows the performance of these defenses against our \sys. It is obvious that these solutions have a minor impact on the prevention of prompt stealing. The main reason is that none of them are specifically designed to prevent prompt-stealing attacks. Therefore, \textit{their modification to the showcase may suffer from low transferability} when attackers use a caption model to extract prompt information. 

\subsection{Dedicated Defense} 
Given the ineffectiveness of the above defenses, we propose a defense dedicated to \sys, dubbed PromptGuard. It aims to preserve the utility of showcasing images for benign users while preventing the reconstruction of prompts by malicious users. The core insight of PromptGuard is to add an invisible perturbation $\epsilon$ to the showcase, which can \textit{disturb the feature in the embedding spaces} of the caption models ($\mathcal{B}$) and modifier matching model ($\mathcal{C}$). A defender can utilize any captioning model and modifier matching model to disrupt the embeddings. Its optimization goal can be formulated as:
\begin{equation}
\begin{aligned}
\min \ & \alpha \cdot \mathcal{D}(\mathcal{B}(s),\mathcal{B}(\hat{s}))+\beta \cdot \mathcal{D}(\mathcal{C}(s),\mathcal{C}(\hat{s})), \\
\text{s.t.} \ (\hat{s}-s) & \in [-\epsilon,\epsilon],
\end{aligned}
\label{function: defense}
\end{equation}
where $\mathcal{D}$ is a distance metric used to measure the feature similarity. Here we choose the cosine similarity, where a larger value indicates a higher similarity and vice versa. Therefore, when using the gradient descent algorithm (\textit{e.g.}, Adam optimizer in our implementation) to optimize Formula~\ref{function: defense}, the feature similarity between $s$ and $\hat{s}$ becomes much lower. The detailed similarity decline process can be found in Figure~\ref{fig: defense-loss}. Specifically, we use the BIM algorithm~\cite{BIM} to generate image perturbation and set the optimization step at 200. We set the perturbation threshold $\epsilon$ to 8/255, a rather small value, to \textit{avoid influencing the normal usage of showcases}. Both the hyperparameters $\alpha$ and $\beta$ are set to 0.5. 

The evaluation results of PromptGuard are presented in Table~\ref{tab: potential defense-1}. The performance of \sys significantly declines under PromptGuard's protection, outperforming conventional defenses by a large margin. This effectiveness stems from PromptGuard's ability to \textit{disrupt showcase embeddings}, reducing their feature similarity to the original ones. As a result, when a captioning model or modifier matching model utilizes these disturbed embeddings for prompt stealing, the generated outputs deviate significantly from the correct semantics. These findings validate the effectiveness of PromptGuard in mitigating prompt-stealing attacks.

\begin{table}[t!]
\caption{Potential defenses against \sys. Here, the arrows indicate the defense effect. }
\vspace{-0.1in}
\label{tab: potential defense-1}
\resizebox*{\linewidth}{!}{
\begin{tabular}{c|cccc}
\toprule
\textbf{Defenses}&${CLIP}_{img}(\downarrow)$&\textbf{$LPIPS(\uparrow)$}&\textbf{$SBERT(\downarrow)$}&\rev{\textbf{$ASR(\downarrow)$}}  \cr \midrule
No Defense&0.901&0.625&0.799&62.5\%  \cr \midrule
Random Noise&0.887 (-0.014)&\textbf{0.775 (+0.150)}&0.795 (-0.004)&58.3\%(-4.2\%)  \\
Puzzle&0.882 (-0.019)&0.659 (+0.034)&0.799 (-0.000)&61.1\%(-1.4\%) \\
Watermark&0.829 (-0.072)&0.735 (+0.110)&0.788 (-0.011)&41.7\%(-20.8\%) \\
\rowcolor{lightgray}PromptGuard&\textbf{0.824 (-0.077)}&0.664 (+0.039)&\textbf{0.774 (-0.025)}&27.8\%(-34.7\%) \cr \midrule
\rowcolor{lightgray}$\sys^*$&0.915 (+0.014)&0.597 (-0.028)&0.804 (+0.005)&67.9\%(+5.4\%) \cr \bottomrule
\end{tabular}
}
\vspace{-0.1in}
\end{table}

\para{Limitation of PromptGuard.}
The success of PromptGuard heavily relies on the assumption that the defender knows the caption models ($\mathcal{B}$) and modifier matching model ($\mathcal{C}$) used by the attacker. However, this assumption does not hold water as the attacker can try other models when he realizes that the chosen $\mathcal{B}$ and $\mathcal{C}$ are shielded. Unfortunately, in this case, the perturbation generated has low transferability to the different caption models and modifier matching models~\cite{huang2019enhancing, wu2018understanding}. This significantly restricts the practical value of PromptGuard. For example, PromptGuard generates the perturbation using BLIP with ViT-base, while the attacker can adopt BLIP with a different structure (\textit{e.g.}, ViT-large) for prompt stealing. The defense results for this setting are shown in Table~\ref{tab: potential defense-1} (\textbf{denoted as $\sys^*$}). We observe that with a different caption model, the stealing performance rises again, even outperforming the no-defense setting. This reveals the resistance of \sys to PromptGuard and the urgent need for more effective defenses. More discussions about such attacks can be found in Appendix~\ref{appendix: defense}. We will discuss more effective defenses in future work. 

\begin{mdframed}[nobreak=true]
\textit{Take-away}: \sys is robust against conventional defenses and resistant to dedicated defense. 
\end{mdframed}

\section{Potential Limitations and Future Directions}

\para{Time Complexity.} The primary limitation of \sys lies in its runtime. Unlike prior works that rely on fussy data collection and pre-training, \sys introduces a lightweight online stealing framework, featuring dynamic modifier generation and a \texttt{Proxy}-integrated search process that is both more adaptive and cost-efficient. On average, \sys takes about two minutes to steal a single prompt. Given that protected showcases are typically curated works, such as artistic prompts, which are relatively few compared to web-scale images, this time complexity would not become a practical obstruction for large-scale attacks. Moreover, as studied in Section~\ref{sec: ablation study}, our parameter choosing is redundant, so that the running time can be further optimized without noticeably affecting performance. Future work can focus on developing more efficient search strategies while preserving effectiveness.

\para{Search Strategy.} This paper proposes a greedy search-based framework for effective prompt stealing, overcoming the overfitting limitations of the traditional training–prediction pipeline. We acknowledge that there may exist interdependence between modifiers so that they have a stronger combined effect than greedy search. However, we argue that, as our candidate modifiers are sampled offline with contextual matching before online search, which ensures every candidate already possesses a reasonably good descriptive capability for the showcase, the improvement of interdependent modifiers is likely to be marginal. However, doing so would significantly increase the computational cost. For example, in our setting with 200 candidates, even considering only pairwise combinations would theoretically slow down the search by approximately 100 times (i.e., 199/2). Given that the potential gain is limited, such a computational cost is unacceptable to a large extent. Future work could explore alternative search strategies beyond our adopted greedy search to further enhance attack performance while preserving effectiveness.

\section{Conclusion}
We propose \sys, an effective, training-free, and search-based prompt-stealing attack against text-to-image generative models. Our main insight is leveraging the \texttt{Proxy} feedback to optimize the prompt to be semantically and perceptually close to the target. To improve the query efficiency and prompt fidelity, we propose the concept of dynamic modifiers, which capture the details and position relations, and utilize the zero-shot ability of BLIP, as well as NLP analysis, to extract such information. We also propose a contextual matching mechanism, which provides a rough ranking order, to improve the greedy search efficacy. We evaluate \sys against large-scale on-sold prompts from PromptBase and AIFrog, which are designed for victim models like Midjourney and DALL$\cdot$E, by querying the well-performed open-source proxy models, including FLUX, ShuttleDiffusion, and SD-3.5. The results demonstrate that \sys achieves an excellent prompt-stealing effect and is resistant to potential defenses.

\section*{Acknowledgment}
We thank all the reviewers for their constructive suggestions that helped improve this paper. We are also grateful to Boheng Li (Nanyang Technological University) for the insightful discussions held during the project. This research/project is supported by the National Research Foundation, Singapore, under its AI Singapore Programme (AISG Award No: AISG3-RPGV-2025-019 and No.AISG4-GC-2023-008-1B), and the Ministry of Education of Singapore under its Academic Research Fund Tier 2 (Grant No: MOE-T2EP20121-0008).

\section{Ethics Considerations and Compliance with the Open Science Policy}

\subsection{Ethics Considerations}
\para{Stakeholder Perspectives and Considerations. }
\begin{icompact}
    \item [1)] Prompt Engineer. Prompt-stealing attacks may directly harm the prompt engineer's commercial interest. First, an attacker can avoid paying for the prompt but steal it. Second, after stealing, the attacker can reload it to prompt markets to violate the original seller's IP further. 
    \item [2)] Prompt Market. Prompt-stealing attacks may harm the prompt markets' commercial interest. First, stealing a prompt instead of buying would lead to no commissions earned for the prompt market. Second, if the attacker loads the stolen prompt to another platform, it would attract potential users away. 
    \item [3)] Artist and Society. Prompt-stealing attacks can extract the style from an image. Therefore, if an artist posts his/her masterpiece, the attacker could extract a prompt that describes their work precisely. Then the attacker can create infinite images that are similar to the artist's works. 
\end{icompact}
\para{Respect for Persons. }
\begin{icompact}
    \item [1)] Notice. We wrote the consent document that details the intended benefits of research activities and the risks to research subjects. 
    \item [2)] Comprehension. The language level is kept to 8th grade or lower to improve the ability of subjects to comprehend the benefits and risks. 
    \item [3)] Voluntariness. The consent document stresses that participation is voluntary and that subjects are free to withdraw from research participation without negative consequences.
\end{icompact}

\para{Positive Impact of Research}
\begin{icompact}
    \item [1)] Identification of Potential Benefits and Harms. The potential harm of prompt stealing is violating the intellectual property of prompt creators. The benefit is by revealing the vulnerability of the prompt, the safety issue of the prompt can attract more people's attention. 
    \item [2)] Mitigation of Realized Harms. We consider preempting the escalation of realized harms by notifying affected parties or otherwise engaging in mitigation actions. More specifically, before our research, we made responsible disclosure and notified all the prompt creators. In addition, we developed a defense against prompt stealing attacks to mitigate the realized harm. 
\end{icompact}

\para{IRB Exemption and Volunteer Protection.}
\begin{icompact}
    \item [1)] \textbf{IRB Exemption.} In our user study, all images used were non-explicit, stylized paintings representing everyday visual content and contained no inappropriate material (\textit{e.g.}, nudity or sexual themes). Accordingly, under standard ethical guidelines, this study qualified for exemption from formal IRB review, consistent with the ``exempt review'' category defined in U.S. IRB protocols (45 CFR 46) because it posed minimal risk to participants. 
    \item [2)] \textbf{Volunteer Protection.} To further safeguard participants, we implemented multiple protective measures. Before participation, all volunteers were informed that the study concerned human judgments of image similarity; the specific research purpose was deliberately described in general terms to prevent priming or potential misuse. Participants were briefed on the procedure and data handling, explicitly reminded that participation was voluntary and could be discontinued at any time, especially if they experienced any form of discomfort, and informed consent was obtained. No personally identifiable information was collected. Besides, all questionnaires were completed offline under direct supervision. Participants were allowed to withdraw immediately if they felt any discomfort, and their status was continuously monitored to ensure safety. Upon completion, participants retained the right to withdraw their responses, and we verified their physical and mental well-being before concluding the study.
\end{icompact}

\subsection{Compliance with the Open Science Policy}
We are committed to adhering to open science policies by sharing the outcomes of our research in an open-access format.
This includes sharing datasets, test cases, scripts, and source code related to our research paper, to promote a broader commitment to open science principles.

\para{Open Sharing of Code and Data. }We will make all artifacts involved in our research, including datasets, test cases, scripts, and source code, publicly available on the GitHub platform to support academic exchange and technological advancement. The models used in our experiments (Stable Diffusion 3.5, FLUX, ShuttleDiffusion) are all open-source and can be accessed and downloaded from HuggingFace. Additionally, the dataset used in this study is the HF-Prompt dataset, an open and non-sensitive dataset available on the HuggingFace platform, ensuring transparency and openness. 

\para{Reproducibility and Replicability. }All artifacts necessary for reproducing our research findings will be made public. We will also provide detailed experimental records and documentation on the GitHub platform, including information on the experimental environment setup, dependencies, and parameter settings, to enable researchers to replicate our experiments.
\bibliographystyle{plain}
\bibliography{sections/9-reference}

\appendices
\section{Method}
\subsection{Problem Motivation}
\label{appendix: motivation}
We specify the motivation for studying prompt stealing. 
\para{Prompts' Value.} The value of \textit{well-crafted prompts} can be validated from two perspectives. Firstly, there are numerous emerging prompt marketplaces and registered prompt engineers. The active prompt transactions prove the economic significance of exquisite prompts. Second, prompt engineering has become a recognized profession. For instance, Adobe offers prompt engineer positions with salaries between USD 162,000 and 301,200~\cite{adobe}, and similar listings appear on Indeed~\cite{indeed}. These trends highlight that \textit{well-crafted prompts} are valuable intellectual assets. 

\para{Attack Scenario.} In some scenarios, users may have vague ideas of what they want to create but struggle to express them effectively, as well-designed prompts embed subtle linguistic cues that significantly influence output quality. Previous work~\cite{oppenlaender2025prompting} validates the ``articulation barrier'' when describing aspects like style or composition, especially for image generation. We target the scenario where users frequently encounter appealing AI-generated artworks online and may seek to reproduce their style, making prompt reconstruction both a realistic and attractive attack scenario.

\subsection{Privacy Protection Standpoint}
\label{appendix: privacy protection standpoint}
From the privacy protection standpoint, prompt-stealing attack poses a significant threat to the confidentiality of prompts. Specifically, in many cases, a user's prompts may interact with sensitive or private data during the generation process (\textit{e.g.}, rely on information retrieved from private databases or internal documents through in-context learning). Then, the recovered prompts include not only the ordinary user's prompts but also the augmented sensitive information, as both are ultimately reflected in the generated images. For example, professional artists might use a prompt such as 'Generate a portrait in the style of my previous drafts (as shown in the provided examples)'. Stealing the prompt of such a generated portrait image could inadvertently expose the artist’s unique stylistic characteristics (\textit{i.e.}, the privacy in this case) embodied in those unreleased drafts. 

\subsection{Dynamic Modifier Extraction}
\label{appendix: dynamic modifier extraction}
We detail the extraction process for dynamic modifiers in Algorithm~\ref{alg:dynamic modifier}. We first generate multiple descriptions of the showcase using a captioning model. We then apply the NLP toolkit Spacy~\cite{spacy} to each caption to obtain its part-of-speech (PoS) tags and dependency tree (DepTree). These two signals are subsequently used to extract subject details and position information, which together form the dynamic modifiers.

\renewcommand{\algorithmicrequire}{\textbf{Input:}}
\renewcommand{\algorithmicensure}{\textbf{Output:}}
\begin{algorithm}[!t] \footnotesize
    
    \caption{$\mathtt{Dynamic}\ \mathtt{Modifier}\ \mathtt{Extraction}$}
    \begin{algorithmic}[1] 
        \Require caption model BLIP $\mathcal{B}$, showcase $s$, caption quantity $q$, and NLP analysis tool $\texttt{Spacy}$. 
        \Ensure extracted dynamic modifier list $\mathcal{M}_{d}$.
    \State $\mathcal{M}_{d}(d)$ $\gets$ $\emptyset$ \Cmnt{Store subject detail}
    \State $\mathcal{M}_{d}(p)$ $\gets$ $\emptyset$ \Cmnt{Store position information}
    \State $\mathcal{P}$ $\gets$ $\mathcal{B}(s, q)$ \Cmnt{Caption q times}
    \For{$p$ in $\mathcal{P}$}
        \State $\mathtt{POS,DepTree}\gets \mathtt{Spacy}(p)$
        \State $\mathcal{C} \gets \mathtt{Spacy}.chunk(p)$ \Cmnt{Get detail chunks}
        \State $\mathcal{M}_{d}(d) \gets \mathcal{M}_{d}(d)\cup \mathcal{C}$ 
        \For{$\mathtt{token}$ in $p$}
            \If{$\mathtt{POS}(\mathtt{token})$ = ``prep''} \Cmnt{Find prepositions}
                \State $\mathtt{child} \gets \mathtt{DepTree}(\mathtt{token})$
                \State $c \gets \mathtt{child}.join()$
                \State $\mathcal{M}_{d}(p) \gets \mathcal{M}_{d}(p)\cup c$
            \EndIf
        \EndFor

    \EndFor

    \State \textbf{return} $\mathcal{M}_{d}(d)$ and $\mathcal{M}_{d}(p)$
    \end{algorithmic}
    \label{alg:dynamic modifier}
\end{algorithm}

\section{Experiment}
\subsection{VLMasExpert}
\label{appendix: more results}
As presented in Table~\ref{tab: main result} and Figure~\ref{fig: visualization-1}, the baseline VLMasExpert, which serves as a potential prompt-stealing attack that considers integrating a human expert (it utilizes VLMs as human experts) into the stealing process, performs worse than \sys. The potential reason can be twofold. First, the same to PromptStealer~\cite{PromptStealing}, the well-performed VLM GPT-4o is trained on specific image-text pairs and may face an over-fitting problem when adapted to the prompt-stealing task. Second, out of copyright protection, most large models, including GPT-4o, are aligned not to answer copyrighted content. Thus, VLMasExpert replaces all the protected subjects with more generalized subjects, which results in less accurate prompt stealing. 

To study which factor accounts for the most of VLM's defect, we experiment by ignoring the test examples that contain copyrighted content, which is denoted as VLMasExpert*. As shown in Table~\ref{tab: VLM alignment}, after disregarding copyrighted showcases, the performance of VLMasExpert is improved, but slightly. This indicates that alignment indeed mitigates VLM's potential for being adopted in the prompt-stealing task. However, the improvement is relatively minor, suggesting that the primary issue is not copyright protection, but rather overfitting. 

\begin{table}[ht!]
\centering
\caption{VLMasExpert without alignment. }
\label{tab: VLM alignment}
\resizebox*{\linewidth}{!}{
\begin{tabular}{c|cccc}
\toprule
\textbf{Method}&\textbf{${CLIP}_{img}(\uparrow)$}&\textbf{$LPIPS(\downarrow)$}&\textbf{$SBERT(\uparrow)$}&\textbf{$ASR(\uparrow)$}  \cr \midrule
VLMasExpert&0.848&0.695&0.776&27.1\%  \\
VLMasExpert*&0.852&0.692&0.785&29.2\%  \\
\rowcolor{lightgray}Prometheus (Ours)&\textbf{0.901}&\textbf{0.653}&\textbf{0.814}&\textbf{58.3\%} \cr \bottomrule
\end{tabular}
}
\end{table}

\subsection{User Study}
\label{appendix: user study}

We specify the user study setting as follows:

\begin{figure*}[t!]
  \centering
  \includegraphics[width=0.9\linewidth]{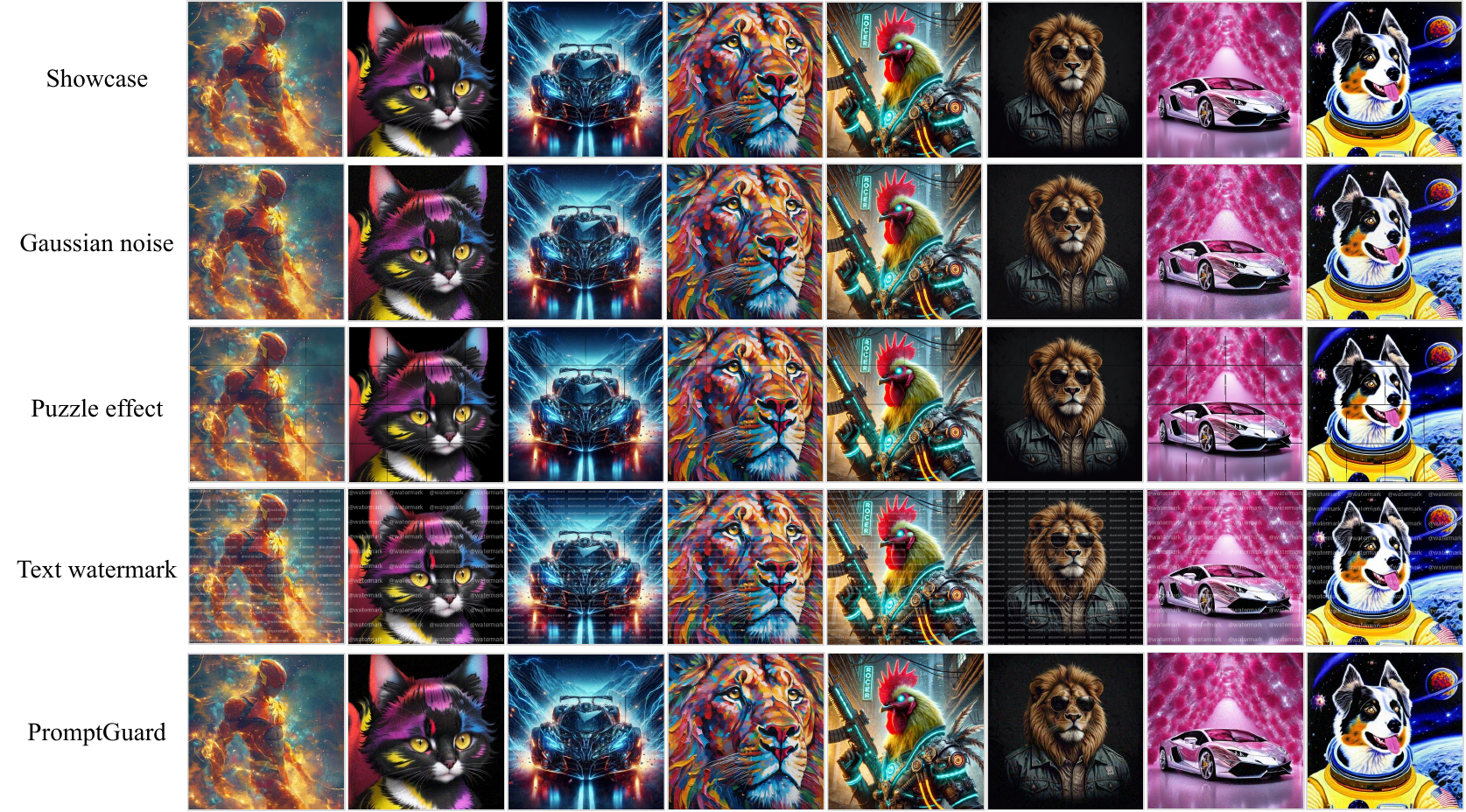} 
  \vspace{-0.1in}
  \caption{Example of generated adversarial example with different defenses. Please zoom out for details. }
  \label{fig: defenses}
  \vspace{-0.1in}
\end{figure*}

\para{Study Detail.} The recruited volunteers were asked to complete a questionnaire on perceptual image similarity. Each participant viewed 12 randomly selected groups of images (a showcase and its recovered versions); each group contained 6 image pairs (5 baselines and our method), yielding 72 questions in total. All images were non-explicit, stylistic paintings within the range of everyday visual content (the examples can be seen in Figure~\ref{fig: visualization-1}). For each question, volunteers rated the perceptual similarity between the showcase and the recovered image using the same scoring criteria as PromptStealer~\cite{PromptStealing}. To avoid publicizing the prompt-stealing setting, we did not disclose the precise research goal and only told participants they were comparing the similarity between two images. To reduce potential bias, we anonymized the methods for every question and randomized the method order within each group to prevent habitual choices and ensure independence across questions. 

\para{Volunteers.} We totally recruited 103 volunteers and obtained 103 valid questionnaires. To eliminate bias, the volunteers are recruited from various backgrounds, including gender, age, and education. Specifically, among all the respondents, 57.28\% are men while 42.72\% are women. As for age, 41.75\% of the surveyed people were between 18-25, and 45.63\% of those were between 26-30. Including respondents aged between 31-40 (total proportion of 10.68\%), these three parts account for 98.06\% of the total. Of the remaining 1.94\%, both the people aged under 18 and those between 51-60 account for 0.97\%. We also realize the effect of education on respondents' preferences. We keep all the questions to 8th grade or lower to ensure they are fully understood by the respondents. Among all the surveyed people, most of them obtained a college degree or above, and only 0.97\% of them have a high school degree. Specifically, 3.88\% of the total respondents have a college degree, and a proportion of 30.1\% have a bachelor's degree. The left part, with a proportion of 65.05\%, occupies the vast majority, and owns postgraduate degree.

\begin{figure*}[t!]
    \centering
    \subfigure[Total Loss]{%
        \includegraphics[width=0.31\linewidth]{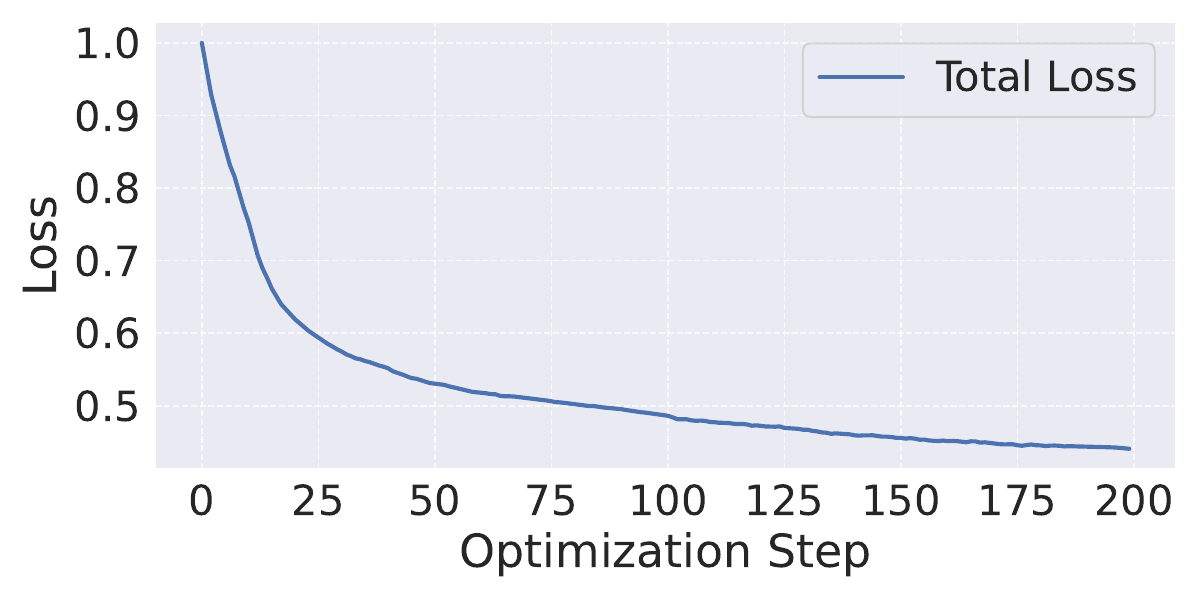}
        \label{fig: defense-total_loss}
    }
    \subfigure[BLIP Loss]{%
        \includegraphics[width=0.31\linewidth]{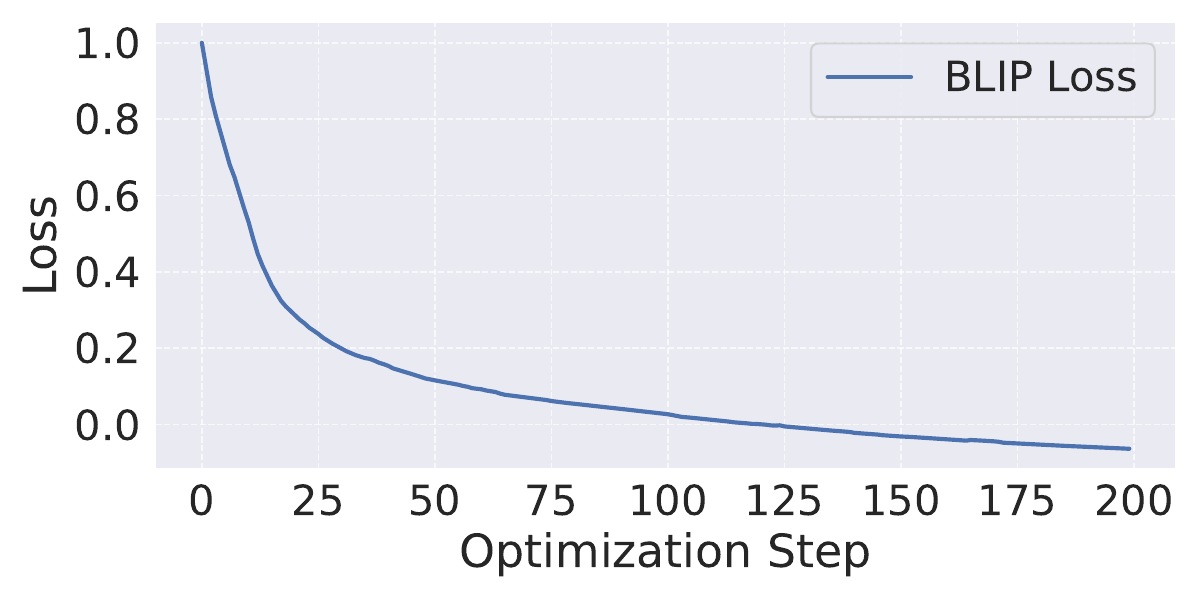}
        \label{fig: defense-blip_loss}
    }
    \subfigure[CLIP Loss]{%
        \includegraphics[width=0.31\linewidth]{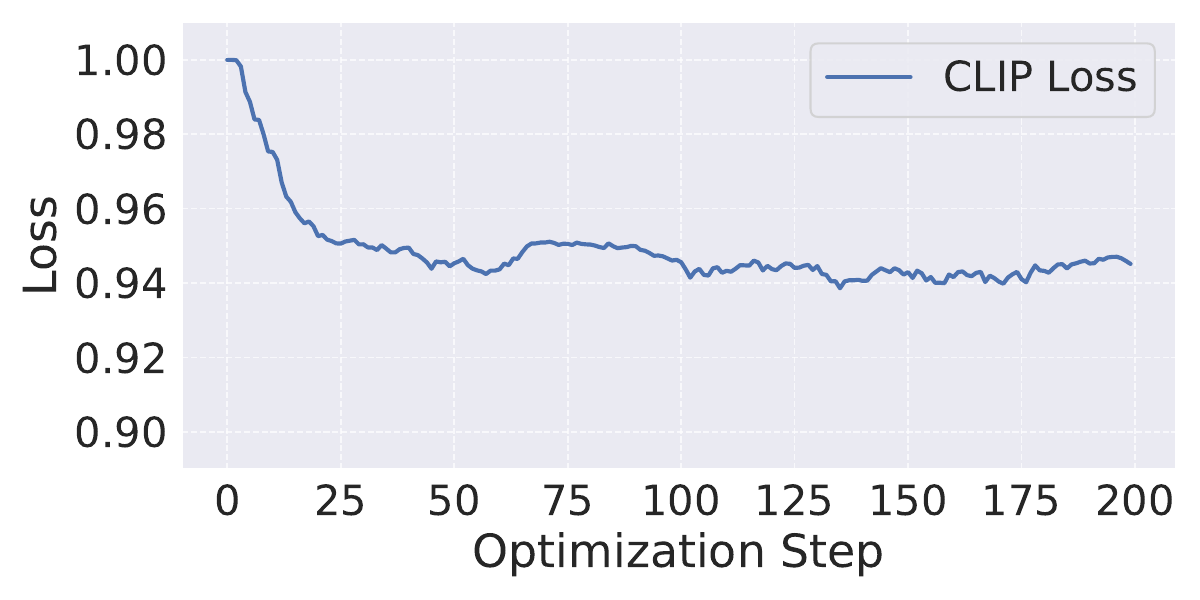}
        \label{fig: defense-clip_loss}
    }
    \vspace{-0.1in}
    \caption{Loss with optimization step. }
    \vspace{-0.1in}
    \label{fig: defense-loss}
\end{figure*}

\subsection{Defense}
\label{appendix: defense}

\para{PromptShield.} PromptShield seeks to defend against prompt-stealing by turning showcases into adversarial examples using gradients from the attacker’s modifier-predictor. It sets the ground-truth prompt without artist modifiers as the adversarial target and then optimizes the showcase with the predictor’s gradients. This approach has four shortcomings. First, it assumes the attacker trains a modifier classifier, which our attack does not, making PromptShield inapplicable and limiting its real-world utility. Second, it assumes white-box access to the attacker’s predictor, which is often unrealistic. Third, it presumes all marketplace prompts include artist modifiers so a target adversarial prompt can be constructed; however, many do not. For example, the prompt in Figure~\ref{fig: showcase} from PromptBase lacks any artist modifier. Finally, even when a prompt contains an artist modifier, removing that element safeguards only the artist modifier while leaving other modifiers unprotected.

\para{Detail of Defenses.} We set the hyperparameter of conventional defenses following the principle that the perturbation should not affect human perception of the subjects and styles. Specifically, for random noise, we generate Gaussian noise with a mean of 0 and a standard deviation of 25, which is almost perceptible to humans. For the puzzle effect, we add a 4×4 grid to the showcase, with variability set to 3. We generate a text watermark ``@watermark'' as the intellectual property claim. The text font size is set to 20, and the gaps in the rows and columns are set to 20 and 30, respectively. We evaluate the defense performance on the RealPrompt dataset to simulate the preventive effect in the real world. 

\para{Analysis of Adaptive Attack.} The proposed defense, named PromptGuard, employs the gradient of a unified loss function to generate adversarial perturbations. The loss curve is shown in Figure~\ref{fig: defense-loss}, where the unified loss ultimately converges around 0.45, demonstrating the defense's effectiveness. However, our observations reveal that the primary decline in the loss comes from BLIP, while CLIP contributes minimally. This can be attributed to CLIP's pre-training on LAION~\cite{laion5b}, a dataset containing billions of examples, which makes it inherently robust to adversarial perturbations. In contrast, BLIP is more fragile due to its significantly smaller training dataset. Based on these findings, when designing adaptive attacks, we focus on modifying the BLIP structure, as adversarial examples exhibit poor transferability across different BLIP models. The core of defending against prompt-stealing attacks is to break the connection between the showcase and its caption, independent of the specific captioning model. Therefore, future defenses should prioritize mitigation strategies that remain robust across different attack model architectures. 

\subsection{More Analysis}
\label{appendix: more analysis}
\para{Main results.} We observe a performance gap between DALLEPrompt and RealPrompt. After reviewing both datasets, we attribute this to two factors: (1) Model type. RealPrompt is largely sourced from commercial T2I systems (\textit{e.g.}, Midjourney, DALL$\cdot$E) with stronger instruction-following, yielding closer prompt–showcase alignment; as a result, showcases more faithfully reflect prompt properties and improve prompt reconstruction accuracy. (2) Prompt type. RealPrompt contains expert-crafted, commercially sold prompts designed by prompt engineers, which tend to be more specific and stylistically distinctive; their showcases are therefore easier to differentiate. By contrast, DALLEPrompt primarily comprises public user prompts that are more generic and can be expressed in many ways, weakening the correspondence between prompts and showcases.

\section{More Related Works}
\subsection{Related Works on Large Language Models}
\label{appendix: llm}

There have been several prompt-stealing attacks against large language models. Yang et al.~\cite{yang2025prsa} propose a prompt attention algorithm to infer the detailed intent of prompts, which realizes accurate prompt stealing even from a single input-output pair. Hui et al.~\cite{hui2024pleak} propose to use a gradient-based method for optimizing the adversarial prompt to steal the system prompt. Sha et al.~\cite{sha2024prompt} introduce a parameter extractor and a prompt reconstructor to extract the prompt, and Tan et al.~\cite{tan2025effectiveness} propose a text gradient-based method to steal prompts in the wild iteratively. Nguyen et al.~\cite{nguyen2025three} propose to adaptively stack ciphers to jailbreak large reasoning models. There are also efforts to protect the privacy of LLMs. For example, Hao et al.~\cite{hao2022iron} initiate the study of private inference on Transformer-based models, and propose the first systematic ZK proof framework for non-linear mathematical functions using the perspective of table lookup~\cite{hao2024scalable}. 

\subsection{Related Works on T2I Diffusion Models}
\label{appendix: t2i}

\para{Attacks}. Given the popularity and great commercial value of T2I models and prompts, lots of work has been dedicated to studying their vulnerability. Wu et al.~\cite{membership-1} use a prompt, captioned with a query image, to conduct a membership inference attack on black-box T2I models. Following this, several works also study the risk of membership leakage~\cite{membership-2, membership-3}. Carlini et al.~\cite{ExtractingData} propose a data extraction attack against T2I models via querying the training prompt. Lukas et al.~\cite{BackdoorTextEncoder} inject backdoors into the text encoder for predefined image generation. Huang et al.~\cite{HuangPersonalizationBackdoor} utilize personalization strategies (a trigger in the prompt) as a shortcut to realize a few-shot backdoor, and Liang et al.~\cite{liang2024badclip} propose a dual-embedding guided backdoor attack against multimodal contrastive learning. Then Liang et al.~\cite{liang2025revisiting} propose a multimodal attribution backdoor attack to improve the attack generalizability. Recent research focuses on the jailbreak vulnerability~\cite{MilliereAdversarial}, where attackers intentionally modify the prompt, inducing the model to generate offensive content even when the safety filter is applied~\cite{FilterJail-1, FilterJail-2, jailbreak1}. For instance, Natalie et al.~\cite{MausAdversarial} propose the concept of adversarial prompts and design a black-box framework for attacking image generation models. Li et al.~\cite{li2025dream} propose a scalable red teaming method via distribution modeling for text-to-image generative systems. Zhao et al.~\cite{zhao2025inception} explore the vulnerability of the memory mechanism and introduce a multiturn jailbreak attack against the text-to-image generation system. Liu et al.~\cite{liu2025t2v} first propose to jailbreak text-to-video models with discrete prompt optimization. 

\para{Defenses}. As for defenses, Wu et al.~\cite{wu2025themis} propose to regulate personalization models in embedding, and Li et al.~\cite{li2025rethinking} propose a comprehensive framework to protect data in the generative AI era. Guo et al.~\cite{guo2025copyrightshield} propose a copyright defense method that integrates poisoned sample detection and adaptive optimization training. Jing et al.~\cite{jing2025promptsafe} propose a text-driven soft prompt-guided defense framework to tackle unsafe generation. Liang et al.~\cite{liang2025t2vshield} introduce a model-agnostic jailbreak defense for text-to-video models. Liang et al.~\cite{liang2025safemobile} first propose a formal and effective autonomous defense system for mobile agents. Zhao et al.~\cite{zhao2025p2p} propose using benign triggers to counter data-poisoning backdoor attacks in LLMs. 

\end{document}